\documentclass[a4paper,twocolumn,aps,prl]{revtex4-2}
\usepackage[english]{babel}
\usepackage[utf8]{inputenc}
\usepackage{amsmath}
\usepackage{amsfonts}
\usepackage{CJK}

\usepackage{xspace}

\usepackage[colorlinks=true,linkcolor=black,citecolor=black]{hyperref}
\usepackage[capitalise]{cleveref}

\usepackage{url}

\usepackage{graphicx}
\graphicspath{ {figures/} }

\usepackage{natbib}

\newcommand{\affA}{Department of Physics and Astronomy, Aarhus University, DK-8000 Aarhus C, Denmark}

\date{\today}
\begin{document}
\begin{CJK*}{UTF8}{gbsn}
	\title{Quantum nonlinear metasurfaces from dual arrays of ultracold atoms}
	
	\author{Simon Panyella Pedersen}
	\email{spp@phys.au.dk}
	\affiliation{\affA}
	\author{Lida Zhang~(\CJKfamily{gbsn}张理达)}
	\email{zhanglida@phys.au.dk}
	\affiliation{\affA}
	\author{Thomas Pohl}
	\email{pohl@phys.au.dk}
	\affiliation{\affA}

\begin{abstract}
Optical interfaces with sub-wavelength patterns make it possible to manipulate light waves beyond the typical capabilities of ordinary optical media. Sub-wavelength arrays of ultracold atoms enable such transformations at very low photon losses. Here, we show how the coupling of light to more than a single atomic array can expand these perspectives into the domain of quantum nonlinear optics. While a single array transmits and reflects light in a highly coherent but largely linear fashion, the combination of two arrays is found to induce strong photon-photon interactions that can convert an incoming classical beam into strongly correlated photonic states. Such quantum metasurfaces open up new possibilities for coherently generating and manipulating nonclassical light, and exploring quantum many-body phenomena in two-dimensional systems of strongly interacting photons.
\end{abstract}

\maketitle
\end{CJK*}

Advances in controlling cold atomic ensembles at the single-particle level \cite{gross_quantum_2017} have enabled the development of novel light-matter interfaces. Recent investigations \cite{zheng_persistent_2013,thompson_coupling_2013,petersen_chiral_2014,tiecke_nanophotonic_2014,goban_superradiance_2015,douglas_quantum_2015,coles_chirality_2016,calajo_atom-field_2016,zoubi_quantum_2017,hamann_nonreciprocity_2018,yu_two-dimensional_2019,ke_inelastic_2019,jones_collectively_2020,bettles_enhanced_2016,facchinetti_storing_2016,shahmoon_cooperative_2017,asenjo-garcia_exponential_2017,ballantine_optical_2020,rui_subradiant_2020,poshakinskiy_quantum_2021,patti_controlling_2021,sheremet_waveguide_2021} have explored various approaches towards strong and coherent coupling to propagating photons, using nanoscale optical waveguides, photonic crystals or regularly arranged atoms in free space. 
In particular, extended two-dimensional lattices of atoms suggest themselves as optical metasurfaces that can be designed and engineered on sub-wavelength scales at the level of individual quantum emitters. Experiments have demonstrated the strong coherent coupling to subradiant collective excitations of atoms in optical lattices \cite{rui_subradiant_2020}, which can enable the near lossless interfacing with a single mode of freely propagating light fields \cite{bettles_enhanced_2016,facchinetti_storing_2016,shahmoon_cooperative_2017,asenjo-garcia_exponential_2017}. This makes it possible to explore the functionalities of optical metasurfaces \cite{Kildishev2013,Yu2014,chen_review_2016,Su2018,Wei2021}, which, e.g., offers new possibilities for highly coherent wavefront engineering \cite{ballantine_optical_2020}. While dense emitter arrangements with an overall sub-wavelength dimension can exhibit high nonlinearities akin to a single atom \cite{cidrim_photon_2020,williamson_superatom_2020}, the extended geometry of two-dimensional surfaces renders such systems intrinsically linear. 
In fact, the simultaneous photon interaction with a large number of atoms, necessary to achieve strong collective coupling, diminishes the otherwise strong optical nonlinearity of each individual quantum emitter, thereby, restricting their collective optical response to the domain of linear optics.

\begin{figure}[t!]
	\centering
	\includegraphics[width=\columnwidth]{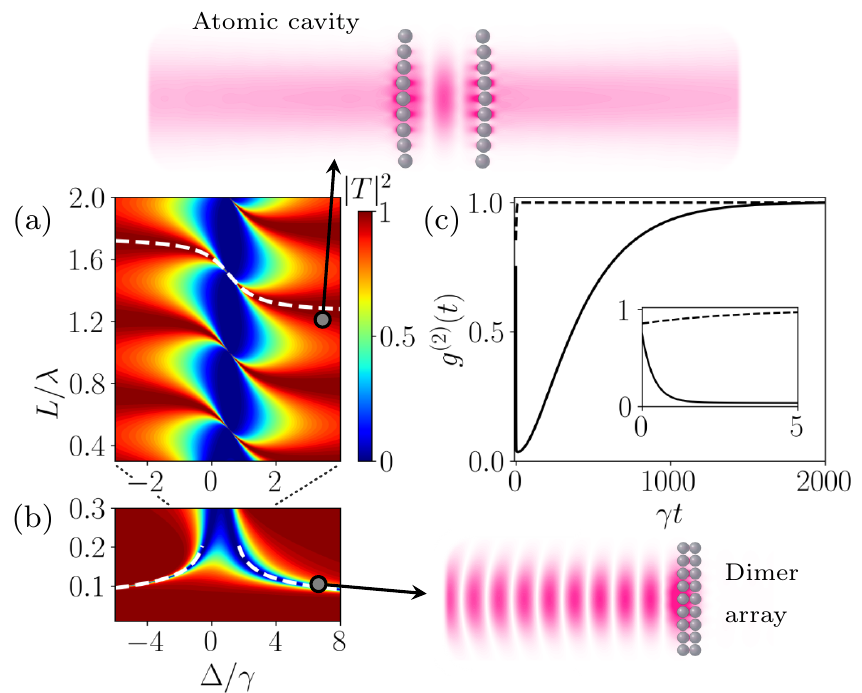}
	\caption{\label{fig1} 
	(a), (b) The transmission spectrum $|T|^2$ of a dual array of two-level emitters exhibits characteristic reflection and transmission resonances as a function of the photon frequency detuning, $\Delta$, and the distance, $L$, between the two two-dimensional atomic lattices.  (b) For small values of $L$ the system behaves as a single layer of superradiant and subradiant atomic dimers that generate two respective reflection resonances, marked by the two white dashed lines. (a) The large-$L$ limit, on the other hand, leads to a series of narrow transmission resonances of the effective atomic resonator, as indicated by the white dashed line in the upper part of the plot [cf. \cref{eq:L}]. (c) In the vicinity of such narrow transmission resonances, the optical response becomes highly nonlinear and generates effective photon-photon interactions that can transform an incident coherent beam into highly nonclassical light, as demonstrated by the depicted two-photon correlation function $ g^{(2)}(t) $ of the transmitted light (solid line). In contrast, light reflected from a single identical array remains largely uncorrelated (dashed line).  The chosen lattice spacing is $a=0.6\lambda$. The transmission in (a) is obtained for infinitely extended arrays, while the calculations in (c) are performed for finite arrays with $ 9\times 9 $ atoms driven by a Gaussian laser beam with a waist of $ w = 1.5\lambda $, $ \Delta = 0.472\gamma $ and $ L = 1.55\lambda $.
	}
\end{figure}

Here, we describe and analyse how strong optical nonlinearities can be obtained in metasurfaces composed of sub-wavelength atomic lattices. Specifically, we will discuss how combining two atomic arrays, which separately are only weakly nonlinear, can greatly enhance their combined optical nonlinearity to a degree that acts on the level of individual photons. The quantum optical nonlinearity in this system arises from narrow transmission resonances around which photons are strongly confined between the two arrays, forming a high finesse optical resonator. We show that such dual-array settings can effectively transform an incident classical beam into strongly correlated states of light, while the statistics of incident photons is left virtually unchanged by each individual array (cf. \cref{fig1}c). The nonlinearity can be traced back to the emergence of an effective photon-photon interaction in the two-dimensional plane of the surface, which is shown to generate in-plane collisions between individual photons. This suggests a promising approach to employing atomic lattices as quantum nonlinear metasurfaces for generating and manipulating nonclassical states of light, and exploring quantum many-body physics with photons \cite{carusotto_quantum_2013,noh_quantum_2017}.

Let us first consider a single two-dimensional square lattice of atoms with a lattice spacing $a$, and which we assume to be infinitely extended in the $x$-$y$ plane. We focus the discussion on two-level systems that are resonantly driven by a coherent cw-field with an electric field amplitude $E_{\rm in}$ and spatial mode function $f({\bf R})$ \footnote{This can be realized \cite{rui_subradiant_2020} by applying a sufficiently strong magnetic field to ensure that only one atomic transition is near-resonant with the incident driving field of a given polarization, and excitation exchange on other dipole transitions is energetically suppressed.}. At the small lattice spacings considered here, exchange of photons across the array generates strong atomic interactions that can be efficiently described within input-output theory by integrating out the photonic degrees of freedom and using a Born-Markov approximation \cite{lehmberg_radiation_1970,gross_superradiance_1982}. This yields a master equation $\partial_t \hat{\rho}=-i[\hat{H},\hat{\rho}]+\mathcal{L}[\hat{\rho}]$ (with $ \hbar = 1 $) for the density matrix, $\hat{\rho}$, of the atomic lattice, where the Hamiltonian and Lindblad operator 
\begin{subequations}
	\begin{align}
	\begin{split}
		\hat{H} =& -\Delta\sum_{n}\hat{\sigma}_{n}^{\dagger}\hat{\sigma}_{n} - \sum_{n}(\Omega_n\hat{\sigma}_{n}^{\dagger}+\Omega_n^{*}\hat{\sigma}_{n}) \\
		& - \sum_{n \neq m}J_{nm}\hat{\sigma}_{n}^{\dagger}\hat{\sigma}_{m} ,
	\end{split} \label{eq:H}\\
	\mathcal{L}[\hat{\rho}] =& \sum_{n,m}\Gamma_{nm}\left(2\hat{\sigma}_{n}\hat{\rho}\hat{\sigma}_{m}^{\dagger} - \left\{\hat{\sigma}_{n}^{\dagger}\hat{\sigma}_{m},\hat{\rho}\right\}\right), \label{eq:Li}
\end{align}
\end{subequations}
describe the exchange of excitations and corresponding collective decay processes due to the photon-mediated dipole-dipole interactions between the atoms \cite{asenjo-garcia_atom-light_2017}.
Here, $\sigma_{n} = | g_{n} \rangle\langle e_{n} |$ denotes the transition operator between the ground state, $| g_{n} \rangle$, and excited state $| e_{n} \rangle$ of an atom at position ${\bf R}_{n}$ in the lattice. The interaction coefficients $J_{nm}$ and decay rates $\Gamma_{nm}$ for two atoms at positions ${\bf R}_n$ and ${\bf R}_m$ are determined by the Green's function tensor of the free-space electromagnetic field \cite{asenjo-garcia_atom-light_2017,suppl}. The atomic transition is driven by the incident light with a frequency detuning $\Delta$ and a single-atom Rabi frequency $\Omega_n=d E_{\rm in} f({\bf R}_n)$, which is determined by the driving-field amplitude and the transition dipole moment $d$ of the two-level emitters. From the solution for $\hat{\rho}$, one can reconstruct the electromagnetic field generated by the driven atomic dipoles. Choosing $f({\bf R}_n)$ as the detection mode for the transmitted light, one obtains \cite{asenjo-garcia_atom-light_2017,suppl}
\begin{equation}\label{eq:E}
	\hat{E} = E_{\rm in} + i\frac{3\pi \gamma}{k^2\eta d}\sum_{n}f^{*}({\bf r}_n)\hat{\sigma}_{n}
\end{equation}
for the electric-field amplitude, $\hat{E}$, of the detected photons, where $k=2\pi/\lambda$ is the wave number of the incident light, $\lambda$ denotes its wavelength, $\gamma = \Gamma_{nn}$ is the decay rate of the individual atoms, and $\eta=\int{\rm d}^2r |f({\bf R})|^2$ with ${\bf r}=(x,y)$. 

For weak plane-wave driving with $f\sim e^{ikz}$, \cref{eq:H,eq:Li,eq:E} yield simple expressions for the transmission and reflection spectra \cite{shahmoon_cooperative_2017}
\begin{equation}\label{eq:tr}
	t=\frac{\langle\hat{E}\rangle}{E_{\rm in}}\:,\quad r=t-1=-\frac{i\tilde{\Gamma}}{\Delta-\tilde{\Delta}+i\tilde{\Gamma}},
\end{equation}
that feature a Lorentzian resonance at the collective Lamb shift $\tilde\Delta=-\sum_{n\neq 0} J_{n0}$ with a width $\tilde{\Gamma} = 3\pi\gamma/k^2a^2$. On resonance, the single atomic layer, thus, reflects incoming photons with unit efficiency and no losses from the incident mode, $|t|^2+|r|^2=1$.

Such high reflectivities are attainable already for remarkably small systems \cite{bettles_enhanced_2016}. For example, a $9\times9$ atomic array with $a=0.6\lambda$ can reflect into the incident mode of a focused Gaussian beam with a waist of $w=1.5\lambda$ with a large reflection amplitude of $|r|=0.998$. On the other hand, the nonlinear response is  very small as the beam still covers a sizeable number of atoms, which substantially diminishes saturation effects. This is seen directly from the second order correlation function $g^{(2)}(t)$ of the reflected light, shown by the dashed line in \cref{fig1}b. Here, one finds only a marginal suppression of simultaneous two-photon reflection, indicating that the reflected light largely retains the classical coherent-state nature of the incident beam ($g^{(2)}\sim1$).

This situation changes dramatically as we add a second atomic array. \cref{fig1}a shows the transmission coefficient $|T|^2$ for a dual-array configuration of two parallel atomic lattices as a function of the detuning $\Delta$ and the distance $L$ between the two arrays. The calculations reveal a series of narrow transmission resonances that extends towards large values of $L$ and a pair of sharp reflection resonances at small array distances. Both regimes can be traced back to the photon-mediated interactions between the two arrays. 

For small distances $L$, atoms in different arrays interact via a dipole-dipole coupling that scales as $J_L\approx-3\gamma/2(kL)^3$ \cite{suppl}.
We can, thus, define symmetric and antisymmetric superposition states, $|\pm\rangle_n$, of a single excitation that is  symmetrically ($|+\rangle_n$) and antisymmetrically ($|-\rangle_n$) shared between two adjacent atoms at a given lattice site $n$. The atomic interaction shifts their respective energies by $\pm J_L$. Therefore, the two atomic dimer states become energetically isolated for small $L$ and separately generate reflection resonances at the collective energies $\tilde{\Delta}_{\pm} \sim \pm L^{-3}$, as indicated by the dashed lines in \cref{fig1}a \cite{suppl}. Their respective widths are given by $\tilde\Gamma_\pm=\tilde\Gamma[1 \pm \cos(kL)]$, such that one finds an ultra-narrow reflection resonance with $\tilde\Gamma_-\ll\tilde\Gamma$, generated by an effective array of subradiant atomic dimer states as $L$ decreases.

For larger values of $L$, the evanescent-field coupling vanishes and the interaction between the arrays is  predominantly generated by propagating photons. The coupling strength $J_L\approx 3\gamma\cos(kL)/kL$, therefore, acquires an oscillating behaviour from the propagation phase and leads to an energy difference $\tilde{\Delta}_{+}-\tilde{\Delta}_{-}\sim\sin(kL)$ that varies periodically with the array distance $L$. As both collective dimer states are excited by the parity-breaking incident field, their interference leads to a series of narrow transmission resonances, akin to electromagnetically induced transparency in three-level systems \cite{fleischhauer_electromagnetically_2005,suppl}. One can obtain a simple expression for the dual-array transmission amplitude \cite{suppl}
\begin{equation}\label{eq:T}
	T = \frac{t^2}{1 - r^2e^{2ikL}} 
\end{equation}
in terms of the reflection and transmission amplitudes, $r$ and $t$, of the individual arrays. Substituting their explicit form, \cref{eq:tr}, we obtain the following condition for perfect transmission ($|T|=1$)
\begin{equation}\label{eq:L}
\Delta - \tilde{\Delta}=-\tilde{\Gamma}	\tan(kL) ,
\end{equation}
that defines the series of transmission resonances. Along these resonances, transmitted photons can acquire a substantial group delay with a delay time \cite{suppl}
\begin{equation}\label{eq:tau}
	\tau = \frac{\tilde{2\Gamma}}{(\Delta - \tilde{\Delta})^2},
\end{equation}
that diverges for resonant detunings $\Delta=\tilde{\Delta}$. The width of the transmission resonances decreases as $\sim1/\tau$ around these values and, therefore, vanishes at $ kL = n\pi $ for integer $ n>0 $. In between the resonances ($ kL = (n+1/2)\pi $), the system features high reflection and behaves largely linear, which can be used to store several delocalized excitations across distant arrays at $L\gg\lambda$ \cite{guimond_subradiant_2019}. 

\cref{eq:T} describes the transmission of a Fabry-P\'{e}rot resonator composed of two identical mirrors with respective reflection and transmission amplitudes $r$ and $t$. 
Here, however, the single-particle saturation of each emitter within the atomic mirrors together with the narrow subradiant transmission resonances can enhance optical nonlinearities and generate exceedingly strong photon-photon interactions.

\begin{figure}[t!]
	\centering
	\includegraphics[width=\columnwidth]{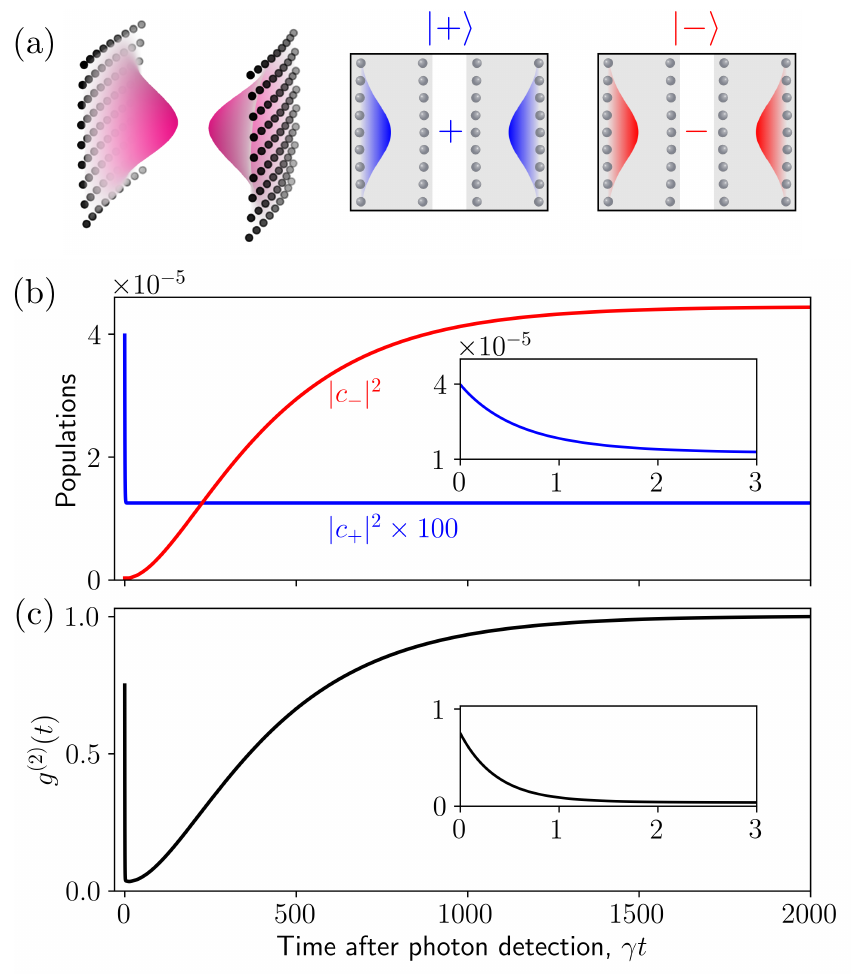}
	\caption{\label{fig2} (a) The characteristic time dependence of the two-photon correlation function, $g^{(2)}(t)$, can be understood by the dynamics of the short-lived ($|+\rangle$) and long-lived ($|-\rangle$) single atomic excitation that is symmetrically ($|+\rangle$) and antisymmetrically ($|-\rangle$) delocalized between the two arrays. (b) Following the detection of a photon, the subsequent population dynamics, $|c_+|^2$ (blue) and $|c_-|^2$ (red), of these two states  agrees with the characteristic time dependence of $g^{(2)}(t)$ shown in panel (c) (see text for more details). The parameters are the same as in \cref{fig1}c.}
\end{figure}

We have studied the signatures of such photon-photon interactions via quantum trajectory wave function simulations \cite{molmer_monte_1993} of the atomic master equation with \cref{eq:H,eq:Li} for finite arrays. Working with finite arrays and focused driving beams, generally entails photon losses that tend to broaden the otherwise ultra-narrow transmission resonances. These effects can be mitigated through a proper choice of the atomic lattices, matching the wavefront profile of the incident beam \cite{guimond_subradiant_2019,suppl}. In fact, already rather small lattices of $9\times9$ atoms permit to generate narrow transmission resonances with linewidths of $\sim10^{-2}\gamma$ and high peak transmission of $|T|\sim0.98$.

\begin{figure}[t]
	\centering
	\includegraphics[width=\columnwidth]{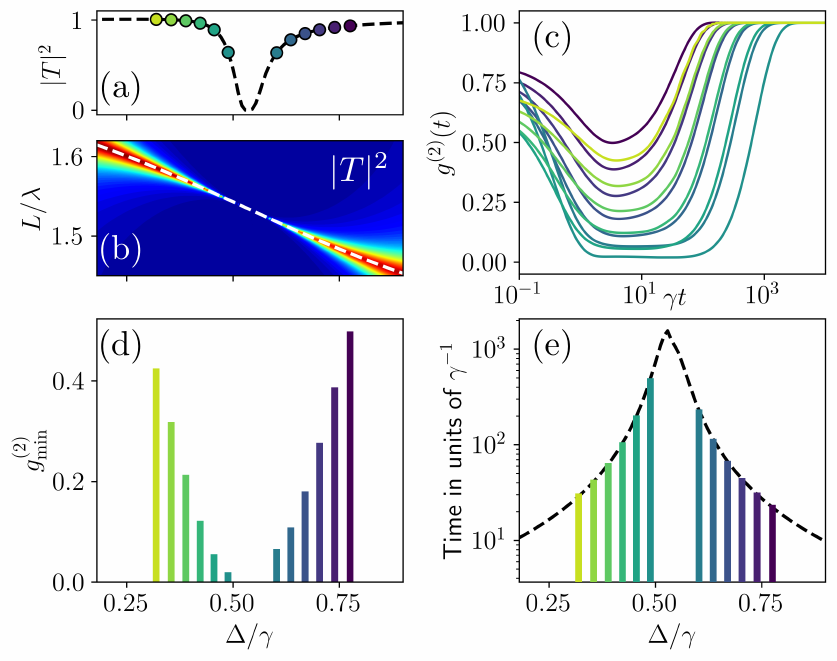}
	\caption{\label{fig3} (a) Transmission of a $ 9\times 9 $ dual array with the same parameters as in \cref{fig1}a along the transmission resonance marked by the white dashed line \cref{fig1}a and panel (b). (c) Two-photon correlation function, $ g^{(2)}(t) $, at the different indicated points along this transmission resonance, as indicated by the color coding. Panel (d) shows the minimum value $ g^{(2)}_{\rm min}={\rm min}_t g^{(2)}(t) $. The color bars in panel (e) show the long timescale on which the correlation functions eventually approach unity, which matches the photon delay time, or photon confinement time, $\tau$. Its asymptotic value for infinite arrays is given by \cref{eq:tau}, while the dashed line in panel (e) has been obtained numerically for the $9\times 9$ arrays.}
\end{figure}

\cref{fig1}c shows the calculated second order correlation function
\begin{equation}\label{eq:g2}
g^{(2)}(t)=\frac{\langle\hat{E}^\dagger(t')\hat{E}^\dagger(t'+t)\hat{E}(t'+t)\hat{E}(t')\rangle}{\langle\hat{E}^\dagger(t')\hat{E}(t')\rangle^2}
\end{equation}
of the transmitted light for an incident cw field with a Gaussian transverse beam profile whose waist is centred right in between the two arrays. We consider long times $t'\rightarrow\infty$, such that \cref{eq:g2} yields the temporal photon-photon correlation in the steady state. Its dependence on the time delay, $t$, between consecutively detected photons indicates the generation of highly nonclassical light. Interestingly, one finds a rapid initial drop of the two-photon correlation function to small values $g^{(2)}\sim0$, which extends over a broad range of delay times between two transmitted photons. These characteristic temporal correlations can be understood as follows. Let us denote the steady state of the two arrays as $|\psi\rangle$. Detection of a transmitted photon in the steady state then projects this state onto $|\bar\psi\rangle=\hat{E}|\psi\rangle/\sqrt{\langle\psi|\hat{E}^{\dagger}\hat{E}|\psi\rangle}$. The correlation function can thus be obtained as 
\begin{equation}
	g^{(2)}(t) = \frac{\langle\bar{\psi}(t'+t)|\hat{E}^{\dagger}\hat{E}|\bar{\psi}(t'+t)\rangle}{\langle\psi(t')|\hat{E}^{\dagger}\hat{E}|\psi(t')\rangle} \label{eq:g2wf}
\end{equation}
from the time evolved state $ |\bar{\psi}(t'+t)\rangle $ following detection of a photon at time $t'$. For weak driving this state is predominantly determined by the collective ground state, $|0\rangle$, and the single-excitation manifold. It can hence be expressed as a superposition, $|\bar{\psi}\rangle = c_{0}|0\rangle + c_{+}| + \rangle + c_{-}| - \rangle $, of the collective superradiant ($|+\rangle$) and subradiant ($|-\rangle$) states of a single atomic excitation that is shared (anti)symmetrically  across the two arrays, as discussed above. Indeed, the time evolution of the populations $|c_\pm|^2$ resembles the temporal photon correlations, see \cref{fig2}. The initial drop of $g^{(2)}$ thus reflects the fast decay of the superradiant excitation on a short timescale $\tau_+$, leaving the dual array depleted of excitations. Its subsequent slow rise, on the other hand, can be traced back to the repopulation of the long-lived subradiant cavity state, $|-\rangle$, on a long timescale $\tau_-$ given by the inverse width of the transmission resonance, which corresponds to the photon delay time, discussed above (cf. \cref{fig3}e and see \cite{suppl}).

Consequently, we expect stronger effects of photon-photon interactions around more narrow transmission lines. This is demonstrated in \cref{fig3}, where we show the two-photon correlation function $g^{(2)}$ while scanning the frequency detuning, $\Delta$, and the array distance, $L$, along the transmission maximum of one of the resonances, as illustrated in \cref{fig3}b. Indeed, we find that the minimum value, $g^{(2)}_{\rm min}={\rm min}_t g^{(2)}(t) $, of the two-photon correlation function decreases as the resonance becomes more narrow and the delay time increases. 
Owing to the finite size of the array, the linear transmission decreases as we approach this regime by varying $L$ (\cref{fig3}a). The associated losses tend to broaden the transmission lines and, as shown in \cref{fig3}e, lead to a maximum delay time of $\tau\sim1000/\gamma$, instead of the otherwise diverging behaviour, discussed above for infinite arrays and plane wave driving  [cf. \cref{eq:tau}]. Larger arrays yield longer photon confinement times, $\tau$, for a given transmission maximum, which, therefore, enhances both the temporal extend and the strength of photon-photon correlations. Remarkably, however, one can reach large single-photon nonlinearities and highly nonclassical light with $g^{(2)}_{\rm min}\sim0$ under conditions of high photon transmission already for moderate system sizes, which are achieved in ongoing optical lattice experiments \cite{gross_quantum_2017,rui_subradiant_2020}.

\begin{figure}[t]
	\centering
	\includegraphics[width=\columnwidth]{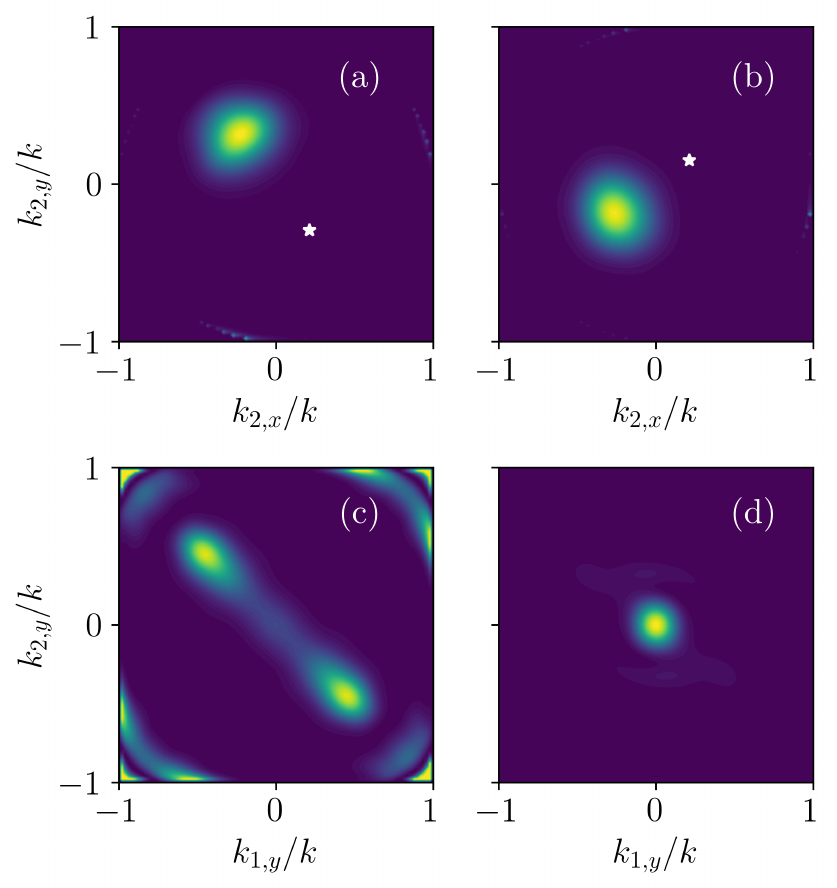}
	\caption{\label{fig4} (a) and (b): Momentum-space two-photon density $ \tilde{\rho}({\bf k}_{1}, {\bf k}_{2},0) $ for $ {\bf k}_{1}\lambda = (1.33, -1.84),\ (1.33 , 0.95) $ respectively. The large density near $ {\bf k}_{1} + {\bf k}_{2} \simeq 0 $ shows the photons have scattered off each other with conserved momentum. (c) and (d): $ \tilde{\rho}({\bf k}_{1}, {\bf k}_{2},t) $ for $ k_{1,x} = k_{2,x} = 0 $ with a time $ t = 0 $ and $ t = 10/\gamma $, respectively, between the detection of the two photons. The time $ t = 10/\gamma $ corresponds to $ g^{(2)}(t) \sim 0 $. Again, the anti-diagonal bar in panel (c) indicates a momentum-conserving scattering among photons. After a time $ t = 10/\gamma $ this correlation has vanished (see text for details). The colour scheme is chosen such that each panel simply shows the qualitative structure. The system parameters are the same as in \cref{fig1}c.}
\end{figure}

We can gain further insights into the generated nonlinearity by considering the two-photon momentum density 
\begin{equation}
	\tilde{\rho}({\bf k}_{1}, {\bf k}_{2},t) = \langle \tilde{E}^{\dagger}({\bf k}_{1},t')\tilde{E}^{\dagger}({\bf k}_{2},t' + t)\tilde{E}({\bf k}_{2},t' + t)\tilde{E}({\bf k}_{1},t') \rangle \label{eq:density}
\end{equation}
in the steady state ($t'\rightarrow\infty$), where $ \tilde{E}({\bf k}_{\perp}) $ is the transverse Fourier transform of the electric field operator of the transmitted light \cite{suppl}. \cref{fig4} shows the in-plane momentum distribution of the two photons.
In \cref{fig4}a and b, we have fixed the transverse momentum ${\bf k}_{1}$ of one photon at the value indicated by the white star and show the equal-time momentum distribution of the other ($t=0$). The distribution of ${\bf k}_{2}$ is sharply peaked around $-{\bf k}_{1}$ indicating that the optical nonlinearity can indeed be understood in terms of effective photon-photon collisions that preserve the total transverse momentum $ {\bf k}_{1} + {\bf k}_{2} \simeq 0 $ of the incident beam. Note that this approximate conservation law already emerges for moderate system sizes of $9\times9$ atoms.
\cref{fig4}c displays the equal-time momentum correlations between the $y$-components of both photons for $k_{1,x}=k_{2,x}=0$. The sharp maximum around $k_{2,y}=-k_{1,y}$ once more reflects the conservation of total momentum, while the variation along the anti-diagonal results from the momentum dependence of the scattering process, i.e., the momentum dependence of the effective photon-photon interaction. The signal in the corners are due to Bragg scattering of the interacting photons. 

\cref{fig4}d shows the same momentum density as in \cref{fig4}c, but for a finite delay time $t=10/\gamma$. In this case we observe practically no correlations in the transverse momenta and the signal corresponds to the transverse mode of the incident beam. This behaviour can be readily understood from the conditioned dynamics of the collective single excitation, discussed above (cf. \cref{fig2}). At small delay times, the two-photon signal naturally stems from the superradiant excitation following detection of the first photon. While the narrow linear transmission line arises from long-lived subradiant excitations, the superradiant mode can only be populated by the interaction between multiple excitations. One therefore finds strong momentum correlations in the transverse two-photon signal, as shown in \cref{fig4}a-c for a vanishing delay time $t=0$. For longer delay times beyond the superradiant lifetime, the superradiant state is depleted and the second photon predominantly stems from the subradiant mode that is repopulated by the continuous optical driving of the arrays. The two detected photons, thus, originate from cascaded excitation and emission processes such that we find negligible  transverse-momentum correlations albeit strong temporal correlations between the two photons. Despite the local nature of the underlying nonlinearity of each individual atom, this mechanism makes it possible to generate strongly correlated photons without significant transverse mode mixing in the plane of the atomic arrays.

Sub-wavelength lattices of ultracold atoms provide a promising platform for the coherent manipulation of optical fields, and in this work we have described how these perspectives can be extended into the domain of quantum nonlinear optics by using two atomic arrays. While large optical nonlinearities can also be generated in atomic ensembles via strong  interactions between high-lying atomic Rydberg states \cite{moreno-cardoner_quantum_2021,zhang_photon-photon_2022,solomons_multi-channel_2021,peyronel_quantum_2012,paris-mandoki_free-space_2017,Stolz2022,Vaneecloo2022}, the present setting yields an alternative mechanism beyond the physics of excitation-blockaded superatoms. 
This is made possible by ultra-narrow transmission resonances that emerge from interference between collective superradiant and subradiant states of the dual array, bearing analogies to electromagnetically induced transparency \cite{fleischhauer_electromagnetically_2005} and the physics of Fano resonators \cite{limonov_fano_2017}. We have demonstrated that the single-photon saturation of each individual atom can generate a strong and finite-ranged effective interaction between photons. Such emerging photon-photon interactions suggest a number of question for future work. We have identified a regime in which strong temporal photon correlations emerge under conditions of very low transverse-mode mixing, thus generating large nonlinearities for freely propagating single photonic modes at greatly suppressed losses. This motivates future explorations of applications as nonlinear quantum optical elements to generating and processing photonic quantum states \cite{witthaut_photon_2012,ralph_photon_2015,Yang2022}, or to study the physics of propagating multi-photon quantum states \cite{mahmoodian_strongly_2018,prasad_correlating_2020,mahmoodian_dynamics_2020,iversen_strongly_2021,iversen_self-ordering_2022} through a many of such nonlinear elements. 

In this work, we have mainly focused on analysing temporal correlations of photons in single transverse modes, drawing analogies to waveguide QED settings \cite{sheremet_waveguide_2021}. In addition, however, the multi-mode physics of large planar arrays should yield an interesting framework for exploring the many-body physics of multiple photons in the two-dimensional plane of the dual-array resonator, and motivates future work on the potential formation and nonlinear dynamics of effective cavity polaritons \cite{carusotto_quantum_2013}. Hereby, our results indicate that this should make it possible to reach the quantum regime of strong interactions between individual polaritons.

We thank Kristian Knakkergaard, Aur\'{e}lien Dantan, and Peter Rabl for valuable discussions. This work was supported by the Carlsberg Foundation through the ''Semper Ardens'' Research Project QCooL, and by the DNRF through the Center of Excellence ''CCQ'' (Grant agreement no.: DNRF156).


\begin{thebibliography}{56}%
	\makeatletter
	\providecommand \@ifxundefined [1]{%
		\@ifx{#1\undefined}
	}%
	\providecommand \@ifnum [1]{%
		\ifnum #1\expandafter \@firstoftwo
		\else \expandafter \@secondoftwo
		\fi
	}%
	\providecommand \@ifx [1]{%
		\ifx #1\expandafter \@firstoftwo
		\else \expandafter \@secondoftwo
		\fi
	}%
	\providecommand \natexlab [1]{#1}%
	\providecommand \enquote  [1]{``#1''}%
	\providecommand \bibnamefont  [1]{#1}%
	\providecommand \bibfnamefont [1]{#1}%
	\providecommand \citenamefont [1]{#1}%
	\providecommand \href@noop [0]{\@secondoftwo}%
	\providecommand \href [0]{\begingroup \@sanitize@url \@href}%
	\providecommand \@href[1]{\@@startlink{#1}\@@href}%
	\providecommand \@@href[1]{\endgroup#1\@@endlink}%
	\providecommand \@sanitize@url [0]{\catcode `\\12\catcode `\$12\catcode
		`\&12\catcode `\#12\catcode `\^12\catcode `\_12\catcode `\%12\relax}%
	\providecommand \@@startlink[1]{}%
	\providecommand \@@endlink[0]{}%
	\providecommand \url  [0]{\begingroup\@sanitize@url \@url }%
	\providecommand \@url [1]{\endgroup\@href {#1}{\urlprefix }}%
	\providecommand \urlprefix  [0]{URL }%
	\providecommand \Eprint [0]{\href }%
	\providecommand \doibase [0]{https://doi.org/}%
	\providecommand \selectlanguage [0]{\@gobble}%
	\providecommand \bibinfo  [0]{\@secondoftwo}%
	\providecommand \bibfield  [0]{\@secondoftwo}%
	\providecommand \translation [1]{[#1]}%
	\providecommand \BibitemOpen [0]{}%
	\providecommand \bibitemStop [0]{}%
	\providecommand \bibitemNoStop [0]{.\EOS\space}%
	\providecommand \EOS [0]{\spacefactor3000\relax}%
	\providecommand \BibitemShut  [1]{\csname bibitem#1\endcsname}%
	\let\auto@bib@innerbib\@empty
	\bibitem [{\citenamefont {Gross}\ and\ \citenamefont
		{Bloch}(2017)}]{gross_quantum_2017}%
	\BibitemOpen
	\bibfield  {author} {\bibinfo {author} {\bibfnamefont {C.}~\bibnamefont
			{Gross}}\ and\ \bibinfo {author} {\bibfnamefont {I.}~\bibnamefont {Bloch}},\
	}\bibfield  {title} {\bibinfo {title} {Quantum simulations with ultracold
			atoms in optical lattices},\ }\href {https://doi.org/10.1126/science.aal3837}
	{\bibfield  {journal} {\bibinfo  {journal} {Science}\ }\textbf {\bibinfo
			{volume} {357}},\ \bibinfo {pages} {995} (\bibinfo {year}
		{2017})}\BibitemShut {NoStop}%
	\bibitem [{\citenamefont {Zheng}\ and\ \citenamefont
		{Baranger}(2013)}]{zheng_persistent_2013}%
	\BibitemOpen
	\bibfield  {author} {\bibinfo {author} {\bibfnamefont {H.}~\bibnamefont
			{Zheng}}\ and\ \bibinfo {author} {\bibfnamefont {H.~U.}\ \bibnamefont
			{Baranger}},\ }\bibfield  {title} {\bibinfo {title} {Persistent {{Quantum
					Beats}} and {{Long-Distance Entanglement}} from {{Waveguide-Mediated
					Interactions}}},\ }\href {https://doi.org/10.1103/PhysRevLett.110.113601}
	{\bibfield  {journal} {\bibinfo  {journal} {Physical Review Letters}\
		}\textbf {\bibinfo {volume} {110}},\ \bibinfo {pages} {113601} (\bibinfo
		{year} {2013})}\BibitemShut {NoStop}%
	\bibitem [{\citenamefont {Thompson}\ \emph {et~al.}(2013)\citenamefont
		{Thompson}, \citenamefont {Tiecke}, \citenamefont {{de Leon}}, \citenamefont
		{Feist}, \citenamefont {Akimov}, \citenamefont {Gullans}, \citenamefont
		{Zibrov}, \citenamefont {Vuleti{\'c}},\ and\ \citenamefont
		{Lukin}}]{thompson_coupling_2013}%
	\BibitemOpen
	\bibfield  {author} {\bibinfo {author} {\bibfnamefont {J.~D.}\ \bibnamefont
			{Thompson}}, \bibinfo {author} {\bibfnamefont {T.~G.}\ \bibnamefont
			{Tiecke}}, \bibinfo {author} {\bibfnamefont {N.~P.}\ \bibnamefont {{de
					Leon}}}, \bibinfo {author} {\bibfnamefont {J.}~\bibnamefont {Feist}},
		\bibinfo {author} {\bibfnamefont {A.~V.}\ \bibnamefont {Akimov}}, \bibinfo
		{author} {\bibfnamefont {M.}~\bibnamefont {Gullans}}, \bibinfo {author}
		{\bibfnamefont {A.~S.}\ \bibnamefont {Zibrov}}, \bibinfo {author}
		{\bibfnamefont {V.}~\bibnamefont {Vuleti{\'c}}},\ and\ \bibinfo {author}
		{\bibfnamefont {M.~D.}\ \bibnamefont {Lukin}},\ }\bibfield  {title} {\bibinfo
		{title} {Coupling a {{Single Trapped Atom}} to a {{Nanoscale Optical
					Cavity}}},\ }\href {https://doi.org/10.1126/science.1237125} {\bibfield
		{journal} {\bibinfo  {journal} {Science}\ }\textbf {\bibinfo {volume}
			{340}},\ \bibinfo {pages} {1202} (\bibinfo {year} {2013})}\BibitemShut
	{NoStop}%
	\bibitem [{\citenamefont {Petersen}\ \emph {et~al.}(2014)\citenamefont
		{Petersen}, \citenamefont {Volz},\ and\ \citenamefont
		{Rauschenbeutel}}]{petersen_chiral_2014}%
	\BibitemOpen
	\bibfield  {author} {\bibinfo {author} {\bibfnamefont {J.}~\bibnamefont
			{Petersen}}, \bibinfo {author} {\bibfnamefont {J.}~\bibnamefont {Volz}},\
		and\ \bibinfo {author} {\bibfnamefont {A.}~\bibnamefont {Rauschenbeutel}},\
	}\bibfield  {title} {\bibinfo {title} {Chiral nanophotonic waveguide
			interface based on spin-orbit interaction of light},\ }\href
	{https://doi.org/10.1126/science.1257671} {\bibfield  {journal} {\bibinfo
			{journal} {Science}\ }\textbf {\bibinfo {volume} {346}},\ \bibinfo {pages}
		{67} (\bibinfo {year} {2014})}\BibitemShut {NoStop}%
	\bibitem [{\citenamefont {Tiecke}\ \emph {et~al.}(2014)\citenamefont {Tiecke},
		\citenamefont {Thompson}, \citenamefont {{de Leon}}, \citenamefont {Liu},
		\citenamefont {Vuleti{\'c}},\ and\ \citenamefont
		{Lukin}}]{tiecke_nanophotonic_2014}%
	\BibitemOpen
	\bibfield  {author} {\bibinfo {author} {\bibfnamefont {T.~G.}\ \bibnamefont
			{Tiecke}}, \bibinfo {author} {\bibfnamefont {J.~D.}\ \bibnamefont
			{Thompson}}, \bibinfo {author} {\bibfnamefont {N.~P.}\ \bibnamefont {{de
					Leon}}}, \bibinfo {author} {\bibfnamefont {L.~R.}\ \bibnamefont {Liu}},
		\bibinfo {author} {\bibfnamefont {V.}~\bibnamefont {Vuleti{\'c}}},\ and\
		\bibinfo {author} {\bibfnamefont {M.~D.}\ \bibnamefont {Lukin}},\ }\bibfield
	{title} {\bibinfo {title} {Nanophotonic quantum phase switch with a single
			atom},\ }\href {https://doi.org/10.1038/nature13188} {\bibfield  {journal}
		{\bibinfo  {journal} {Nature}\ }\textbf {\bibinfo {volume} {508}},\ \bibinfo
		{pages} {241} (\bibinfo {year} {2014})}\BibitemShut {NoStop}%
	\bibitem [{\citenamefont {Goban}\ \emph {et~al.}(2015)\citenamefont {Goban},
		\citenamefont {Hung}, \citenamefont {Hood}, \citenamefont {Yu}, \citenamefont
		{Muniz}, \citenamefont {Painter},\ and\ \citenamefont
		{Kimble}}]{goban_superradiance_2015}%
	\BibitemOpen
	\bibfield  {author} {\bibinfo {author} {\bibfnamefont {A.}~\bibnamefont
			{Goban}}, \bibinfo {author} {\bibfnamefont {C.-L.}\ \bibnamefont {Hung}},
		\bibinfo {author} {\bibfnamefont {J.~D.}\ \bibnamefont {Hood}}, \bibinfo
		{author} {\bibfnamefont {S.-P.}\ \bibnamefont {Yu}}, \bibinfo {author}
		{\bibfnamefont {J.~A.}\ \bibnamefont {Muniz}}, \bibinfo {author}
		{\bibfnamefont {O.}~\bibnamefont {Painter}},\ and\ \bibinfo {author}
		{\bibfnamefont {H.~J.}\ \bibnamefont {Kimble}},\ }\bibfield  {title}
	{\bibinfo {title} {Superradiance for {{Atoms Trapped}} along a {{Photonic
					Crystal Waveguide}}},\ }\href
	{https://doi.org/10.1103/PhysRevLett.115.063601} {\bibfield  {journal}
		{\bibinfo  {journal} {Physical Review Letters}\ }\textbf {\bibinfo {volume}
			{115}},\ \bibinfo {pages} {063601} (\bibinfo {year} {2015})}\BibitemShut
	{NoStop}%
	\bibitem [{\citenamefont {Douglas}\ \emph {et~al.}(2015)\citenamefont
		{Douglas}, \citenamefont {Habibian}, \citenamefont {Hung}, \citenamefont
		{Gorshkov}, \citenamefont {Kimble},\ and\ \citenamefont
		{Chang}}]{douglas_quantum_2015}%
	\BibitemOpen
	\bibfield  {author} {\bibinfo {author} {\bibfnamefont {J.~S.}\ \bibnamefont
			{Douglas}}, \bibinfo {author} {\bibfnamefont {H.}~\bibnamefont {Habibian}},
		\bibinfo {author} {\bibfnamefont {C.-L.}\ \bibnamefont {Hung}}, \bibinfo
		{author} {\bibfnamefont {A.~V.}\ \bibnamefont {Gorshkov}}, \bibinfo {author}
		{\bibfnamefont {H.~J.}\ \bibnamefont {Kimble}},\ and\ \bibinfo {author}
		{\bibfnamefont {D.~E.}\ \bibnamefont {Chang}},\ }\bibfield  {title} {\bibinfo
		{title} {Quantum many-body models with cold atoms coupled to photonic
			crystals},\ }\href {https://doi.org/10.1038/nphoton.2015.57} {\bibfield
		{journal} {\bibinfo  {journal} {Nature Photonics}\ }\textbf {\bibinfo
			{volume} {9}},\ \bibinfo {pages} {326} (\bibinfo {year} {2015})},\ \Eprint
	{https://arxiv.org/abs/1312.2435} {arxiv:1312.2435} \BibitemShut {NoStop}%
	\bibitem [{\citenamefont {Coles}\ \emph {et~al.}(2016)\citenamefont {Coles},
		\citenamefont {Price}, \citenamefont {Dixon}, \citenamefont {Royall},
		\citenamefont {Clarke}, \citenamefont {Kok}, \citenamefont {Skolnick},
		\citenamefont {Fox},\ and\ \citenamefont {Makhonin}}]{coles_chirality_2016}%
	\BibitemOpen
	\bibfield  {author} {\bibinfo {author} {\bibfnamefont {R.~J.}\ \bibnamefont
			{Coles}}, \bibinfo {author} {\bibfnamefont {D.~M.}\ \bibnamefont {Price}},
		\bibinfo {author} {\bibfnamefont {J.~E.}\ \bibnamefont {Dixon}}, \bibinfo
		{author} {\bibfnamefont {B.}~\bibnamefont {Royall}}, \bibinfo {author}
		{\bibfnamefont {E.}~\bibnamefont {Clarke}}, \bibinfo {author} {\bibfnamefont
			{P.}~\bibnamefont {Kok}}, \bibinfo {author} {\bibfnamefont {M.~S.}\
			\bibnamefont {Skolnick}}, \bibinfo {author} {\bibfnamefont {A.~M.}\
			\bibnamefont {Fox}},\ and\ \bibinfo {author} {\bibfnamefont {M.~N.}\
			\bibnamefont {Makhonin}},\ }\bibfield  {title} {\bibinfo {title} {Chirality
			of nanophotonic waveguide with embedded quantum emitter for unidirectional
			spin transfer},\ }\href {https://doi.org/10.1038/ncomms11183} {\bibfield
		{journal} {\bibinfo  {journal} {Nature Communications}\ }\textbf {\bibinfo
			{volume} {7}},\ \bibinfo {pages} {11183} (\bibinfo {year}
		{2016})}\BibitemShut {NoStop}%
	\bibitem [{\citenamefont {Calaj{\'o}}\ \emph {et~al.}(2016)\citenamefont
		{Calaj{\'o}}, \citenamefont {Ciccarello}, \citenamefont {Chang},\ and\
		\citenamefont {Rabl}}]{calajo_atom-field_2016}%
	\BibitemOpen
	\bibfield  {author} {\bibinfo {author} {\bibfnamefont {G.}~\bibnamefont
			{Calaj{\'o}}}, \bibinfo {author} {\bibfnamefont {F.}~\bibnamefont
			{Ciccarello}}, \bibinfo {author} {\bibfnamefont {D.}~\bibnamefont {Chang}},\
		and\ \bibinfo {author} {\bibfnamefont {P.}~\bibnamefont {Rabl}},\ }\bibfield
	{title} {\bibinfo {title} {Atom-field dressed states in slow-light waveguide
			{{QED}}},\ }\href {https://doi.org/10.1103/PhysRevA.93.033833} {\bibfield
		{journal} {\bibinfo  {journal} {Physical Review A}\ }\textbf {\bibinfo
			{volume} {93}},\ \bibinfo {pages} {033833} (\bibinfo {year}
		{2016})}\BibitemShut {NoStop}%
	\bibitem [{\citenamefont {Zoubi}\ and\ \citenamefont
		{Hammerer}(2017)}]{zoubi_quantum_2017}%
	\BibitemOpen
	\bibfield  {author} {\bibinfo {author} {\bibfnamefont {H.}~\bibnamefont
			{Zoubi}}\ and\ \bibinfo {author} {\bibfnamefont {K.}~\bibnamefont
			{Hammerer}},\ }\bibfield  {title} {\bibinfo {title} {Quantum {{Nonlinear
					Optics}} in {{Optomechanical Nanoscale Waveguides}}},\ }\href
	{https://doi.org/10.1103/PhysRevLett.119.123602} {\bibfield  {journal}
		{\bibinfo  {journal} {Physical Review Letters}\ }\textbf {\bibinfo {volume}
			{119}},\ \bibinfo {pages} {123602} (\bibinfo {year} {2017})}\BibitemShut
	{NoStop}%
	\bibitem [{\citenamefont {Hamann}\ \emph {et~al.}(2018)\citenamefont {Hamann},
		\citenamefont {M{\"u}ller}, \citenamefont {Jerger}, \citenamefont {Zanner},
		\citenamefont {Combes}, \citenamefont {Pletyukhov}, \citenamefont {Weides},
		\citenamefont {Stace},\ and\ \citenamefont
		{Fedorov}}]{hamann_nonreciprocity_2018}%
	\BibitemOpen
	\bibfield  {author} {\bibinfo {author} {\bibfnamefont {A.~R.}\ \bibnamefont
			{Hamann}}, \bibinfo {author} {\bibfnamefont {C.}~\bibnamefont {M{\"u}ller}},
		\bibinfo {author} {\bibfnamefont {M.}~\bibnamefont {Jerger}}, \bibinfo
		{author} {\bibfnamefont {M.}~\bibnamefont {Zanner}}, \bibinfo {author}
		{\bibfnamefont {J.}~\bibnamefont {Combes}}, \bibinfo {author} {\bibfnamefont
			{M.}~\bibnamefont {Pletyukhov}}, \bibinfo {author} {\bibfnamefont
			{M.}~\bibnamefont {Weides}}, \bibinfo {author} {\bibfnamefont {T.~M.}\
			\bibnamefont {Stace}},\ and\ \bibinfo {author} {\bibfnamefont
			{A.}~\bibnamefont {Fedorov}},\ }\bibfield  {title} {\bibinfo {title}
		{Nonreciprocity realized with quantum nonlinearity},\ }\href
	{https://doi.org/10.1103/PhysRevLett.121.123601} {\bibfield  {journal}
		{\bibinfo  {journal} {Physical Review Letters}\ }\textbf {\bibinfo {volume}
			{121}},\ \bibinfo {pages} {123601} (\bibinfo {year} {2018})},\ \Eprint
	{https://arxiv.org/abs/1806.00182} {arxiv:1806.00182} \BibitemShut {NoStop}%
	\bibitem [{\citenamefont {Yu}\ \emph {et~al.}(2019)\citenamefont {Yu},
		\citenamefont {Muniz}, \citenamefont {Hung},\ and\ \citenamefont
		{Kimble}}]{yu_two-dimensional_2019}%
	\BibitemOpen
	\bibfield  {author} {\bibinfo {author} {\bibfnamefont {S.-P.}\ \bibnamefont
			{Yu}}, \bibinfo {author} {\bibfnamefont {J.~A.}\ \bibnamefont {Muniz}},
		\bibinfo {author} {\bibfnamefont {C.-L.}\ \bibnamefont {Hung}},\ and\
		\bibinfo {author} {\bibfnamefont {H.~J.}\ \bibnamefont {Kimble}},\ }\bibfield
	{title} {\bibinfo {title} {Two-dimensional photonic crystals for engineering
			atom\textendash light interactions},\ }\href
	{https://doi.org/10.1073/pnas.1822110116} {\bibfield  {journal} {\bibinfo
			{journal} {Proceedings of the National Academy of Sciences}\ }\textbf
		{\bibinfo {volume} {116}},\ \bibinfo {pages} {12743} (\bibinfo {year}
		{2019})}\BibitemShut {NoStop}%
	\bibitem [{\citenamefont {Ke}\ \emph {et~al.}(2019)\citenamefont {Ke},
		\citenamefont {Poshakinskiy}, \citenamefont {Lee}, \citenamefont {Kivshar},\
		and\ \citenamefont {Poddubny}}]{ke_inelastic_2019}%
	\BibitemOpen
	\bibfield  {author} {\bibinfo {author} {\bibfnamefont {Y.}~\bibnamefont
			{Ke}}, \bibinfo {author} {\bibfnamefont {A.~V.}\ \bibnamefont
			{Poshakinskiy}}, \bibinfo {author} {\bibfnamefont {C.}~\bibnamefont {Lee}},
		\bibinfo {author} {\bibfnamefont {Y.~S.}\ \bibnamefont {Kivshar}},\ and\
		\bibinfo {author} {\bibfnamefont {A.~N.}\ \bibnamefont {Poddubny}},\
	}\bibfield  {title} {\bibinfo {title} {Inelastic {{Scattering}} of {{Photon
					Pairs}} in {{Qubit Arrays}} with {{Subradiant States}}},\ }\href
	{https://doi.org/10.1103/PhysRevLett.123.253601} {\bibfield  {journal}
		{\bibinfo  {journal} {Physical Review Letters}\ }\textbf {\bibinfo {volume}
			{123}},\ \bibinfo {pages} {253601} (\bibinfo {year} {2019})}\BibitemShut
	{NoStop}%
	\bibitem [{\citenamefont {Jones}\ \emph {et~al.}(2020)\citenamefont {Jones},
		\citenamefont {Buonaiuto}, \citenamefont {Lang}, \citenamefont {Lesanovsky},\
		and\ \citenamefont {Olmos}}]{jones_collectively_2020}%
	\BibitemOpen
	\bibfield  {author} {\bibinfo {author} {\bibfnamefont {R.}~\bibnamefont
			{Jones}}, \bibinfo {author} {\bibfnamefont {G.}~\bibnamefont {Buonaiuto}},
		\bibinfo {author} {\bibfnamefont {B.}~\bibnamefont {Lang}}, \bibinfo {author}
		{\bibfnamefont {I.}~\bibnamefont {Lesanovsky}},\ and\ \bibinfo {author}
		{\bibfnamefont {B.}~\bibnamefont {Olmos}},\ }\bibfield  {title} {\bibinfo
		{title} {Collectively {{Enhanced Chiral Photon Emission}} from an {{Atomic
					Array}} near a {{Nanofiber}}},\ }\href
	{https://doi.org/10.1103/PhysRevLett.124.093601} {\bibfield  {journal}
		{\bibinfo  {journal} {Physical Review Letters}\ }\textbf {\bibinfo {volume}
			{124}},\ \bibinfo {pages} {093601} (\bibinfo {year} {2020})}\BibitemShut
	{NoStop}%
	\bibitem [{\citenamefont {Bettles}\ \emph {et~al.}(2016)\citenamefont
		{Bettles}, \citenamefont {Gardiner},\ and\ \citenamefont
		{Adams}}]{bettles_enhanced_2016}%
	\BibitemOpen
	\bibfield  {author} {\bibinfo {author} {\bibfnamefont {R.~J.}\ \bibnamefont
			{Bettles}}, \bibinfo {author} {\bibfnamefont {S.~A.}\ \bibnamefont
			{Gardiner}},\ and\ \bibinfo {author} {\bibfnamefont {C.~S.}\ \bibnamefont
			{Adams}},\ }\bibfield  {title} {\bibinfo {title} {Enhanced optical cross
			section via collective coupling of atomic dipoles in a {{2D}} array},\ }\href
	{https://doi.org/10.1103/PhysRevLett.116.103602} {\bibfield  {journal}
		{\bibinfo  {journal} {Physical Review Letters}\ }\textbf {\bibinfo {volume}
			{116}},\ \bibinfo {pages} {103602} (\bibinfo {year} {2016})},\ \Eprint
	{https://arxiv.org/abs/1510.07855} {arxiv:1510.07855} \BibitemShut {NoStop}%
	\bibitem [{\citenamefont {Facchinetti}\ \emph {et~al.}(2016)\citenamefont
		{Facchinetti}, \citenamefont {Jenkins},\ and\ \citenamefont
		{Ruostekoski}}]{facchinetti_storing_2016}%
	\BibitemOpen
	\bibfield  {author} {\bibinfo {author} {\bibfnamefont {G.}~\bibnamefont
			{Facchinetti}}, \bibinfo {author} {\bibfnamefont {S.~D.}\ \bibnamefont
			{Jenkins}},\ and\ \bibinfo {author} {\bibfnamefont {J.}~\bibnamefont
			{Ruostekoski}},\ }\bibfield  {title} {\bibinfo {title} {Storing {{Light}}
			with {{Subradiant Correlations}} in {{Arrays}} of {{Atoms}}},\ }\href
	{https://doi.org/10.1103/PhysRevLett.117.243601} {\bibfield  {journal}
		{\bibinfo  {journal} {Physical Review Letters}\ }\textbf {\bibinfo {volume}
			{117}},\ \bibinfo {pages} {243601} (\bibinfo {year} {2016})}\BibitemShut
	{NoStop}%
	\bibitem [{\citenamefont {Shahmoon}\ \emph {et~al.}(2017)\citenamefont
		{Shahmoon}, \citenamefont {Wild}, \citenamefont {Lukin},\ and\ \citenamefont
		{Yelin}}]{shahmoon_cooperative_2017}%
	\BibitemOpen
	\bibfield  {author} {\bibinfo {author} {\bibfnamefont {E.}~\bibnamefont
			{Shahmoon}}, \bibinfo {author} {\bibfnamefont {D.~S.}\ \bibnamefont {Wild}},
		\bibinfo {author} {\bibfnamefont {M.~D.}\ \bibnamefont {Lukin}},\ and\
		\bibinfo {author} {\bibfnamefont {S.~F.}\ \bibnamefont {Yelin}},\ }\bibfield
	{title} {\bibinfo {title} {Cooperative {{Resonances}} in {{Light Scattering}}
			from {{Two-Dimensional Atomic Arrays}}},\ }\href
	{https://doi.org/10.1103/PhysRevLett.118.113601} {\bibfield  {journal}
		{\bibinfo  {journal} {Physical Review Letters}\ }\textbf {\bibinfo {volume}
			{118}},\ \bibinfo {pages} {113601} (\bibinfo {year} {2017})}\BibitemShut
	{NoStop}%
	\bibitem [{\citenamefont {{Asenjo-Garcia}}\ \emph
		{et~al.}(2017{\natexlab{a}})\citenamefont {{Asenjo-Garcia}}, \citenamefont
		{{Moreno-Cardoner}}, \citenamefont {Albrecht}, \citenamefont {Kimble},\ and\
		\citenamefont {Chang}}]{asenjo-garcia_exponential_2017}%
	\BibitemOpen
	\bibfield  {author} {\bibinfo {author} {\bibfnamefont {A.}~\bibnamefont
			{{Asenjo-Garcia}}}, \bibinfo {author} {\bibfnamefont {M.}~\bibnamefont
			{{Moreno-Cardoner}}}, \bibinfo {author} {\bibfnamefont {A.}~\bibnamefont
			{Albrecht}}, \bibinfo {author} {\bibfnamefont {H.~J.}\ \bibnamefont
			{Kimble}},\ and\ \bibinfo {author} {\bibfnamefont {D.~E.}\ \bibnamefont
			{Chang}},\ }\bibfield  {title} {\bibinfo {title} {Exponential improvement in
			photon storage fidelities using subradiance and "selective radiance" in
			atomic arrays},\ }\href {https://doi.org/10.1103/PhysRevX.7.031024}
	{\bibfield  {journal} {\bibinfo  {journal} {Physical Review X}\ }\textbf
		{\bibinfo {volume} {7}},\ \bibinfo {pages} {031024} (\bibinfo {year}
		{2017}{\natexlab{a}})},\ \Eprint {https://arxiv.org/abs/1703.03382}
	{arxiv:1703.03382} \BibitemShut {NoStop}%
	\bibitem [{\citenamefont {Ballantine}\ and\ \citenamefont
		{Ruostekoski}(2020)}]{ballantine_optical_2020}%
	\BibitemOpen
	\bibfield  {author} {\bibinfo {author} {\bibfnamefont {K.~E.}\ \bibnamefont
			{Ballantine}}\ and\ \bibinfo {author} {\bibfnamefont {J.}~\bibnamefont
			{Ruostekoski}},\ }\bibfield  {title} {\bibinfo {title} {Optical {{Magnetism}}
			and {{Huygens}}' {{Surfaces}} in {{Arrays}} of {{Atoms Induced}} by
			{{Cooperative Responses}}},\ }\href
	{https://doi.org/10.1103/PhysRevLett.125.143604} {\bibfield  {journal}
		{\bibinfo  {journal} {Physical Review Letters}\ }\textbf {\bibinfo {volume}
			{125}},\ \bibinfo {pages} {143604} (\bibinfo {year} {2020})},\ \Eprint
	{https://arxiv.org/abs/2002.12930} {arxiv:2002.12930} \BibitemShut {NoStop}%
	\bibitem [{\citenamefont {Rui}\ \emph {et~al.}(2020)\citenamefont {Rui},
		\citenamefont {Wei}, \citenamefont {{Rubio-Abadal}}, \citenamefont
		{Hollerith}, \citenamefont {Zeiher}, \citenamefont {{Stamper-Kurn}},
		\citenamefont {Gross},\ and\ \citenamefont {Bloch}}]{rui_subradiant_2020}%
	\BibitemOpen
	\bibfield  {author} {\bibinfo {author} {\bibfnamefont {J.}~\bibnamefont
			{Rui}}, \bibinfo {author} {\bibfnamefont {D.}~\bibnamefont {Wei}}, \bibinfo
		{author} {\bibfnamefont {A.}~\bibnamefont {{Rubio-Abadal}}}, \bibinfo
		{author} {\bibfnamefont {S.}~\bibnamefont {Hollerith}}, \bibinfo {author}
		{\bibfnamefont {J.}~\bibnamefont {Zeiher}}, \bibinfo {author} {\bibfnamefont
			{D.~M.}\ \bibnamefont {{Stamper-Kurn}}}, \bibinfo {author} {\bibfnamefont
			{C.}~\bibnamefont {Gross}},\ and\ \bibinfo {author} {\bibfnamefont
			{I.}~\bibnamefont {Bloch}},\ }\bibfield  {title} {\bibinfo {title} {A
			subradiant optical mirror formed by a single structured atomic layer},\
	}\href {https://doi.org/10.1038/s41586-020-2463-x} {\bibfield  {journal}
		{\bibinfo  {journal} {Nature}\ }\textbf {\bibinfo {volume} {583}},\ \bibinfo
		{pages} {369} (\bibinfo {year} {2020})},\ \Eprint
	{https://arxiv.org/abs/2001.00795} {arxiv:2001.00795} \BibitemShut {NoStop}%
	\bibitem [{\citenamefont {Poshakinskiy}\ \emph {et~al.}(2021)\citenamefont
		{Poshakinskiy}, \citenamefont {Zhong},\ and\ \citenamefont
		{Poddubny}}]{poshakinskiy_quantum_2021}%
	\BibitemOpen
	\bibfield  {author} {\bibinfo {author} {\bibfnamefont {A.~V.}\ \bibnamefont
			{Poshakinskiy}}, \bibinfo {author} {\bibfnamefont {J.}~\bibnamefont
			{Zhong}},\ and\ \bibinfo {author} {\bibfnamefont {A.~N.}\ \bibnamefont
			{Poddubny}},\ }\bibfield  {title} {\bibinfo {title} {Quantum {{Chaos Driven}}
			by {{Long-Range Waveguide-Mediated Interactions}}},\ }\href
	{https://doi.org/10.1103/PhysRevLett.126.203602} {\bibfield  {journal}
		{\bibinfo  {journal} {Physical Review Letters}\ }\textbf {\bibinfo {volume}
			{126}},\ \bibinfo {pages} {203602} (\bibinfo {year} {2021})}\BibitemShut
	{NoStop}%
	\bibitem [{\citenamefont {Patti}\ \emph {et~al.}(2021)\citenamefont {Patti},
		\citenamefont {Wild}, \citenamefont {Shahmoon}, \citenamefont {Lukin},\ and\
		\citenamefont {Yelin}}]{patti_controlling_2021}%
	\BibitemOpen
	\bibfield  {author} {\bibinfo {author} {\bibfnamefont {T.~L.}\ \bibnamefont
			{Patti}}, \bibinfo {author} {\bibfnamefont {D.~S.}\ \bibnamefont {Wild}},
		\bibinfo {author} {\bibfnamefont {E.}~\bibnamefont {Shahmoon}}, \bibinfo
		{author} {\bibfnamefont {M.~D.}\ \bibnamefont {Lukin}},\ and\ \bibinfo
		{author} {\bibfnamefont {S.~F.}\ \bibnamefont {Yelin}},\ }\bibfield  {title}
	{\bibinfo {title} {Controlling interactions between quantum emitters using
			atom arrays},\ }\href {https://doi.org/10.1103/PhysRevLett.126.223602}
	{\bibfield  {journal} {\bibinfo  {journal} {Physical Review Letters}\
		}\textbf {\bibinfo {volume} {126}},\ \bibinfo {pages} {223602} (\bibinfo
		{year} {2021})},\ \Eprint {https://arxiv.org/abs/2005.03495}
	{arxiv:2005.03495} \BibitemShut {NoStop}%
	\bibitem [{\citenamefont {Sheremet}\ \emph {et~al.}(2021)\citenamefont
		{Sheremet}, \citenamefont {Petrov}, \citenamefont {Iorsh}, \citenamefont
		{Poshakinskiy},\ and\ \citenamefont {Poddubny}}]{sheremet_waveguide_2021}%
	\BibitemOpen
	\bibfield  {author} {\bibinfo {author} {\bibfnamefont {A.~S.}\ \bibnamefont
			{Sheremet}}, \bibinfo {author} {\bibfnamefont {M.~I.}\ \bibnamefont
			{Petrov}}, \bibinfo {author} {\bibfnamefont {I.~V.}\ \bibnamefont {Iorsh}},
		\bibinfo {author} {\bibfnamefont {A.~V.}\ \bibnamefont {Poshakinskiy}},\ and\
		\bibinfo {author} {\bibfnamefont {A.~N.}\ \bibnamefont {Poddubny}},\
	}\bibfield  {title} {\bibinfo {title} {Waveguide quantum electrodynamics:
			Collective radiance and photon-photon correlations},\ }\bibfield  {journal}
	{\bibinfo  {journal} {arXiv:2103.06824 [quant-ph]}\ }\href
	{https://doi.org/10.48550/arXiv.2103.06824} {10.48550/arXiv.2103.06824}
	(\bibinfo {year} {2021}),\ \Eprint {https://arxiv.org/abs/2103.06824}
	{arxiv:2103.06824 [quant-ph]} \BibitemShut {NoStop}%
	\bibitem [{\citenamefont {Kildishev}\ \emph {et~al.}(2013)\citenamefont
		{Kildishev}, \citenamefont {Boltasseva},\ and\ \citenamefont
		{Shalaev}}]{Kildishev2013}%
	\BibitemOpen
	\bibfield  {author} {\bibinfo {author} {\bibfnamefont {A.~V.}\ \bibnamefont
			{Kildishev}}, \bibinfo {author} {\bibfnamefont {A.}~\bibnamefont
			{Boltasseva}},\ and\ \bibinfo {author} {\bibfnamefont {V.~M.}\ \bibnamefont
			{Shalaev}},\ }\bibfield  {title} {\bibinfo {title} {Planar photonics with
			metasurfaces},\ }\href {https://doi.org/10.1126/science.1232009} {\bibfield
		{journal} {\bibinfo  {journal} {Science}\ }\textbf {\bibinfo {volume}
			{339}},\ \bibinfo {pages} {1232009} (\bibinfo {year} {2013})}\BibitemShut
	{NoStop}%
	\bibitem [{\citenamefont {Yu}\ and\ \citenamefont {Capasso}(2014)}]{Yu2014}%
	\BibitemOpen
	\bibfield  {author} {\bibinfo {author} {\bibfnamefont {N.}~\bibnamefont
			{Yu}}\ and\ \bibinfo {author} {\bibfnamefont {F.}~\bibnamefont {Capasso}},\
	}\bibfield  {title} {\bibinfo {title} {Flat optics with designer
			metasurfaces},\ }\href {https://doi.org/10.1038/nmat3839} {\bibfield
		{journal} {\bibinfo  {journal} {Nature Materials}\ }\textbf {\bibinfo
			{volume} {13}},\ \bibinfo {pages} {139} (\bibinfo {year} {2014})}\BibitemShut
	{NoStop}%
	\bibitem [{\citenamefont {Chen}\ \emph {et~al.}(2016)\citenamefont {Chen},
		\citenamefont {Taylor},\ and\ \citenamefont {Yu}}]{chen_review_2016}%
	\BibitemOpen
	\bibfield  {author} {\bibinfo {author} {\bibfnamefont {H.-T.}\ \bibnamefont
			{Chen}}, \bibinfo {author} {\bibfnamefont {A.~J.}\ \bibnamefont {Taylor}},\
		and\ \bibinfo {author} {\bibfnamefont {N.}~\bibnamefont {Yu}},\ }\bibfield
	{title} {\bibinfo {title} {A review of metasurfaces: Physics and
			applications},\ }\href {https://doi.org/10.1088/0034-4885/79/7/076401}
	{\bibfield  {journal} {\bibinfo  {journal} {Reports on Progress in Physics}\
		}\textbf {\bibinfo {volume} {79}},\ \bibinfo {pages} {076401} (\bibinfo
		{year} {2016})}\BibitemShut {NoStop}%
	\bibitem [{\citenamefont {Su}\ \emph {et~al.}(2018)\citenamefont {Su},
		\citenamefont {Chu}, \citenamefont {Sun},\ and\ \citenamefont
		{Tsai}}]{Su2018}%
	\BibitemOpen
	\bibfield  {author} {\bibinfo {author} {\bibfnamefont {V.-C.}\ \bibnamefont
			{Su}}, \bibinfo {author} {\bibfnamefont {C.~H.}\ \bibnamefont {Chu}},
		\bibinfo {author} {\bibfnamefont {G.}~\bibnamefont {Sun}},\ and\ \bibinfo
		{author} {\bibfnamefont {D.~P.}\ \bibnamefont {Tsai}},\ }\bibfield  {title}
	{\bibinfo {title} {Advances in optical metasurfaces: fabrication and
			applications},\ }\href {https://doi.org/10.1364/OE.26.013148} {\bibfield
		{journal} {\bibinfo  {journal} {Opt. Express}\ }\textbf {\bibinfo {volume}
			{26}},\ \bibinfo {pages} {13148} (\bibinfo {year} {2018})}\BibitemShut
	{NoStop}%
	\bibitem [{\citenamefont {Qiu}\ \emph {et~al.}(2021)\citenamefont {Qiu},
		\citenamefont {Zhang}, \citenamefont {Hu},\ and\ \citenamefont
		{Kivshar}}]{Wei2021}%
	\BibitemOpen
	\bibfield  {author} {\bibinfo {author} {\bibfnamefont {C.-W.}\ \bibnamefont
			{Qiu}}, \bibinfo {author} {\bibfnamefont {T.}~\bibnamefont {Zhang}}, \bibinfo
		{author} {\bibfnamefont {G.}~\bibnamefont {Hu}},\ and\ \bibinfo {author}
		{\bibfnamefont {Y.}~\bibnamefont {Kivshar}},\ }\bibfield  {title} {\bibinfo
		{title} {Quo vadis, metasurfaces?},\ }\bibfield  {booktitle} {\emph {\bibinfo
			{booktitle} {Nano Letters}},\ }\href
	{https://doi.org/10.1021/acs.nanolett.1c00828} {\bibfield  {journal}
		{\bibinfo  {journal} {Nano Letters}\ }\textbf {\bibinfo {volume} {21}},\
		\bibinfo {pages} {5461} (\bibinfo {year} {2021})}\BibitemShut {NoStop}%
	\bibitem [{\citenamefont {Cidrim}\ \emph {et~al.}(2020)\citenamefont {Cidrim},
		\citenamefont {{do Espirito Santo}}, \citenamefont {Schachenmayer},
		\citenamefont {Kaiser},\ and\ \citenamefont
		{Bachelard}}]{cidrim_photon_2020}%
	\BibitemOpen
	\bibfield  {author} {\bibinfo {author} {\bibfnamefont {A.}~\bibnamefont
			{Cidrim}}, \bibinfo {author} {\bibfnamefont {T.~S.}\ \bibnamefont {{do
					Espirito Santo}}}, \bibinfo {author} {\bibfnamefont {J.}~\bibnamefont
			{Schachenmayer}}, \bibinfo {author} {\bibfnamefont {R.}~\bibnamefont
			{Kaiser}},\ and\ \bibinfo {author} {\bibfnamefont {R.}~\bibnamefont
			{Bachelard}},\ }\bibfield  {title} {\bibinfo {title} {Photon {{Blockade}}
			with {{Ground-State Neutral Atoms}}},\ }\href
	{https://doi.org/10.1103/PhysRevLett.125.073601} {\bibfield  {journal}
		{\bibinfo  {journal} {Physical Review Letters}\ }\textbf {\bibinfo {volume}
			{125}},\ \bibinfo {pages} {073601} (\bibinfo {year} {2020})}\BibitemShut
	{NoStop}%
	\bibitem [{\citenamefont {Williamson}\ \emph {et~al.}(2020)\citenamefont
		{Williamson}, \citenamefont {Borgh},\ and\ \citenamefont
		{Ruostekoski}}]{williamson_superatom_2020}%
	\BibitemOpen
	\bibfield  {author} {\bibinfo {author} {\bibfnamefont {L.~A.}\ \bibnamefont
			{Williamson}}, \bibinfo {author} {\bibfnamefont {M.~O.}\ \bibnamefont
			{Borgh}},\ and\ \bibinfo {author} {\bibfnamefont {J.}~\bibnamefont
			{Ruostekoski}},\ }\bibfield  {title} {\bibinfo {title} {Superatom {{Picture}}
			of {{Collective Nonclassical Light Emission}} and {{Dipole Blockade}} in
			{{Atom Arrays}}},\ }\href {https://doi.org/10.1103/PhysRevLett.125.073602}
	{\bibfield  {journal} {\bibinfo  {journal} {Physical Review Letters}\
		}\textbf {\bibinfo {volume} {125}},\ \bibinfo {pages} {073602} (\bibinfo
		{year} {2020})}\BibitemShut {NoStop}%
	\bibitem [{\citenamefont {Carusotto}\ and\ \citenamefont
		{Ciuti}(2013)}]{carusotto_quantum_2013}%
	\BibitemOpen
	\bibfield  {author} {\bibinfo {author} {\bibfnamefont {I.}~\bibnamefont
			{Carusotto}}\ and\ \bibinfo {author} {\bibfnamefont {C.}~\bibnamefont
			{Ciuti}},\ }\bibfield  {title} {\bibinfo {title} {Quantum fluids of light},\
	}\href {https://doi.org/10.1103/RevModPhys.85.299} {\bibfield  {journal}
		{\bibinfo  {journal} {Reviews of Modern Physics}\ }\textbf {\bibinfo {volume}
			{85}},\ \bibinfo {pages} {299} (\bibinfo {year} {2013})}\BibitemShut
	{NoStop}%
	\bibitem [{\citenamefont {Noh}\ and\ \citenamefont
		{Angelakis}(2017)}]{noh_quantum_2017}%
	\BibitemOpen
	\bibfield  {author} {\bibinfo {author} {\bibfnamefont {C.}~\bibnamefont
			{Noh}}\ and\ \bibinfo {author} {\bibfnamefont {D.~G.}\ \bibnamefont
			{Angelakis}},\ }\bibfield  {title} {\bibinfo {title} {Quantum simulations and
			many-body physics with light},\ }\href
	{https://doi.org/10.1088/0034-4885/80/1/016401} {\bibfield  {journal}
		{\bibinfo  {journal} {Reports on Progress in Physics}\ }\textbf {\bibinfo
			{volume} {80}},\ \bibinfo {pages} {016401} (\bibinfo {year}
		{2017})}\BibitemShut {NoStop}%
	\bibitem [{Note1()}]{Note1}%
	\BibitemOpen
	\bibinfo {note} {This can be realized \cite {rui_subradiant_2020} by applying
		a sufficiently strong magnetic field to ensure that only one atomic
		transition is near-resonant with the incident driving field of a given
		polarization, and excitation exchange on other dipole transitions is
		energetically suppressed.}\BibitemShut {Stop}%
	\bibitem [{\citenamefont {Lehmberg}(1970)}]{lehmberg_radiation_1970}%
	\BibitemOpen
	\bibfield  {author} {\bibinfo {author} {\bibfnamefont {R.~H.}\ \bibnamefont
			{Lehmberg}},\ }\bibfield  {title} {\bibinfo {title} {Radiation from an {{N}}
			-{{Atom System}}. {{I}}. {{General Formalism}}},\ }\href
	{https://doi.org/10.1103/PhysRevA.2.883} {\bibfield  {journal} {\bibinfo
			{journal} {Physical Review A}\ }\textbf {\bibinfo {volume} {2}},\ \bibinfo
		{pages} {883} (\bibinfo {year} {1970})}\BibitemShut {NoStop}%
	\bibitem [{\citenamefont {Gross}\ and\ \citenamefont
		{Haroche}(1982)}]{gross_superradiance_1982}%
	\BibitemOpen
	\bibfield  {author} {\bibinfo {author} {\bibfnamefont {M.}~\bibnamefont
			{Gross}}\ and\ \bibinfo {author} {\bibfnamefont {S.}~\bibnamefont
			{Haroche}},\ }\bibfield  {title} {\bibinfo {title} {Superradiance: {{An}}
			essay on the theory of collective spontaneous emission},\ }\href
	{https://doi.org/10.1016/0370-1573(82)90102-8} {\bibfield  {journal}
		{\bibinfo  {journal} {Physics Reports}\ }\textbf {\bibinfo {volume} {93}},\
		\bibinfo {pages} {301} (\bibinfo {year} {1982})}\BibitemShut {NoStop}%
	\bibitem [{\citenamefont {{Asenjo-Garcia}}\ \emph
		{et~al.}(2017{\natexlab{b}})\citenamefont {{Asenjo-Garcia}}, \citenamefont
		{Hood}, \citenamefont {Chang},\ and\ \citenamefont
		{Kimble}}]{asenjo-garcia_atom-light_2017}%
	\BibitemOpen
	\bibfield  {author} {\bibinfo {author} {\bibfnamefont {A.}~\bibnamefont
			{{Asenjo-Garcia}}}, \bibinfo {author} {\bibfnamefont {J.~D.}\ \bibnamefont
			{Hood}}, \bibinfo {author} {\bibfnamefont {D.~E.}\ \bibnamefont {Chang}},\
		and\ \bibinfo {author} {\bibfnamefont {H.~J.}\ \bibnamefont {Kimble}},\
	}\bibfield  {title} {\bibinfo {title} {Atom-light interactions in
			quasi-one-dimensional nanostructures: {{A Green}}'s-function perspective},\
	}\href {https://doi.org/10.1103/PhysRevA.95.033818} {\bibfield  {journal}
		{\bibinfo  {journal} {Physical Review A}\ }\textbf {\bibinfo {volume} {95}},\
		\bibinfo {pages} {033818} (\bibinfo {year} {2017}{\natexlab{b}})}\BibitemShut
	{NoStop}%
	\bibitem [{sup()}]{suppl}%
	\BibitemOpen
	\href@noop {} {}\bibinfo {note} {See Supplemental Material for details on the
		linear response of the dual array, the effects of interactions between two
		arrays, group delay of transmitted photons, finite size effects, the
		numerical analysis of subradiant and superradiant states in finite dual
		arrays, and determining the E-field operator of a certain detection mode and
		the Fourier transform of the E-field.}\BibitemShut {Stop}%
	\bibitem [{\citenamefont {Fleischhauer}\ \emph {et~al.}(2005)\citenamefont
		{Fleischhauer}, \citenamefont {Imamoglu},\ and\ \citenamefont
		{Marangos}}]{fleischhauer_electromagnetically_2005}%
	\BibitemOpen
	\bibfield  {author} {\bibinfo {author} {\bibfnamefont {M.}~\bibnamefont
			{Fleischhauer}}, \bibinfo {author} {\bibfnamefont {A.}~\bibnamefont
			{Imamoglu}},\ and\ \bibinfo {author} {\bibfnamefont {J.~P.}\ \bibnamefont
			{Marangos}},\ }\bibfield  {title} {\bibinfo {title} {Electromagnetically
			induced transparency: {{Optics}} in coherent media},\ }\href
	{https://doi.org/10.1103/RevModPhys.77.633} {\bibfield  {journal} {\bibinfo
			{journal} {Reviews of Modern Physics}\ }\textbf {\bibinfo {volume} {77}},\
		\bibinfo {pages} {633} (\bibinfo {year} {2005})}\BibitemShut {NoStop}%
	\bibitem [{\citenamefont {Guimond}\ \emph {et~al.}(2019)\citenamefont
		{Guimond}, \citenamefont {Grankin}, \citenamefont {Vasilyev}, \citenamefont
		{Vermersch},\ and\ \citenamefont {Zoller}}]{guimond_subradiant_2019}%
	\BibitemOpen
	\bibfield  {author} {\bibinfo {author} {\bibfnamefont {P.-O.}\ \bibnamefont
			{Guimond}}, \bibinfo {author} {\bibfnamefont {A.}~\bibnamefont {Grankin}},
		\bibinfo {author} {\bibfnamefont {D.~V.}\ \bibnamefont {Vasilyev}}, \bibinfo
		{author} {\bibfnamefont {B.}~\bibnamefont {Vermersch}},\ and\ \bibinfo
		{author} {\bibfnamefont {P.}~\bibnamefont {Zoller}},\ }\bibfield  {title}
	{\bibinfo {title} {Subradiant {{Bell}} states in distant atomic arrays},\
	}\href {https://doi.org/10.1103/PhysRevLett.122.093601} {\bibfield  {journal}
		{\bibinfo  {journal} {Physical Review Letters}\ }\textbf {\bibinfo {volume}
			{122}},\ \bibinfo {pages} {093601} (\bibinfo {year} {2019})},\ \Eprint
	{https://arxiv.org/abs/1901.02665} {arxiv:1901.02665} \BibitemShut {NoStop}%
	\bibitem [{\citenamefont {M{\o}lmer}\ \emph {et~al.}(1993)\citenamefont
		{M{\o}lmer}, \citenamefont {Castin},\ and\ \citenamefont
		{Dalibard}}]{molmer_monte_1993}%
	\BibitemOpen
	\bibfield  {author} {\bibinfo {author} {\bibfnamefont {K.}~\bibnamefont
			{M{\o}lmer}}, \bibinfo {author} {\bibfnamefont {Y.}~\bibnamefont {Castin}},\
		and\ \bibinfo {author} {\bibfnamefont {J.}~\bibnamefont {Dalibard}},\
	}\bibfield  {title} {\bibinfo {title} {Monte {{Carlo}} wave-function method
			in quantum optics},\ }\href {https://doi.org/10.1364/JOSAB.10.000524}
	{\bibfield  {journal} {\bibinfo  {journal} {Journal of the Optical Society of
				America B}\ }\textbf {\bibinfo {volume} {10}},\ \bibinfo {pages} {524}
		(\bibinfo {year} {1993})}\BibitemShut {NoStop}%
	\bibitem [{\citenamefont {{Moreno-Cardoner}}\ \emph {et~al.}(2021)\citenamefont
		{{Moreno-Cardoner}}, \citenamefont {Goncalves},\ and\ \citenamefont
		{Chang}}]{moreno-cardoner_quantum_2021}%
	\BibitemOpen
	\bibfield  {author} {\bibinfo {author} {\bibfnamefont {M.}~\bibnamefont
			{{Moreno-Cardoner}}}, \bibinfo {author} {\bibfnamefont {D.}~\bibnamefont
			{Goncalves}},\ and\ \bibinfo {author} {\bibfnamefont {D.~E.}\ \bibnamefont
			{Chang}},\ }\bibfield  {title} {\bibinfo {title} {Quantum nonlinear optics
			based on two-dimensional {{Rydberg}} atom arrays},\ }\href@noop {} {\bibfield
		{journal} {\bibinfo  {journal} {arXiv:2101.01936 [quant-ph]}\ } (\bibinfo
		{year} {2021})},\ \Eprint {https://arxiv.org/abs/2101.01936}
	{arxiv:2101.01936 [quant-ph]} \BibitemShut {NoStop}%
	\bibitem [{\citenamefont {Zhang}\ \emph {et~al.}(2022)\citenamefont {Zhang},
		\citenamefont {Walther}, \citenamefont {M{\o}lmer},\ and\ \citenamefont
		{Pohl}}]{zhang_photon-photon_2022}%
	\BibitemOpen
	\bibfield  {author} {\bibinfo {author} {\bibfnamefont {L.}~\bibnamefont
			{Zhang}}, \bibinfo {author} {\bibfnamefont {V.}~\bibnamefont {Walther}},
		\bibinfo {author} {\bibfnamefont {K.}~\bibnamefont {M{\o}lmer}},\ and\
		\bibinfo {author} {\bibfnamefont {T.}~\bibnamefont {Pohl}},\ }\bibfield
	{title} {\bibinfo {title} {Photon-photon interactions in {{Rydberg-atom}}
			arrays},\ }\href {https://doi.org/10.22331/q-2022-03-30-674} {\bibfield
		{journal} {\bibinfo  {journal} {Quantum}\ }\textbf {\bibinfo {volume} {6}},\
		\bibinfo {pages} {674} (\bibinfo {year} {2022})}\BibitemShut {NoStop}%
	\bibitem [{\citenamefont {Solomons}\ and\ \citenamefont
		{Shahmoon}(2021)}]{solomons_multi-channel_2021}%
	\BibitemOpen
	\bibfield  {author} {\bibinfo {author} {\bibfnamefont {Y.}~\bibnamefont
			{Solomons}}\ and\ \bibinfo {author} {\bibfnamefont {E.}~\bibnamefont
			{Shahmoon}},\ }\bibfield  {title} {\bibinfo {title} {Multi-channel waveguide
			{{QED}} with atomic arrays in free space},\ }\href@noop {} {\bibfield
		{journal} {\bibinfo  {journal} {arXiv:2111.11515 [quant-ph]}\ } (\bibinfo
		{year} {2021})},\ \Eprint {https://arxiv.org/abs/2111.11515}
	{arxiv:2111.11515 [quant-ph]} \BibitemShut {NoStop}%
	\bibitem [{\citenamefont {Peyronel}\ \emph {et~al.}(2012)\citenamefont
		{Peyronel}, \citenamefont {Firstenberg}, \citenamefont {Liang}, \citenamefont
		{Hofferberth}, \citenamefont {Gorshkov}, \citenamefont {Pohl}, \citenamefont
		{Lukin},\ and\ \citenamefont {Vuleti{\'c}}}]{peyronel_quantum_2012}%
	\BibitemOpen
	\bibfield  {author} {\bibinfo {author} {\bibfnamefont {T.}~\bibnamefont
			{Peyronel}}, \bibinfo {author} {\bibfnamefont {O.}~\bibnamefont
			{Firstenberg}}, \bibinfo {author} {\bibfnamefont {Q.-Y.}\ \bibnamefont
			{Liang}}, \bibinfo {author} {\bibfnamefont {S.}~\bibnamefont {Hofferberth}},
		\bibinfo {author} {\bibfnamefont {A.~V.}\ \bibnamefont {Gorshkov}}, \bibinfo
		{author} {\bibfnamefont {T.}~\bibnamefont {Pohl}}, \bibinfo {author}
		{\bibfnamefont {M.~D.}\ \bibnamefont {Lukin}},\ and\ \bibinfo {author}
		{\bibfnamefont {V.}~\bibnamefont {Vuleti{\'c}}},\ }\bibfield  {title}
	{\bibinfo {title} {Quantum nonlinear optics with single photons enabled by
			strongly interacting atoms},\ }\href {https://doi.org/10.1038/nature11361}
	{\bibfield  {journal} {\bibinfo  {journal} {Nature}\ }\textbf {\bibinfo
			{volume} {488}},\ \bibinfo {pages} {57} (\bibinfo {year} {2012})}\BibitemShut
	{NoStop}%
	\bibitem [{\citenamefont {{Paris-Mandoki}}\ \emph {et~al.}(2017)\citenamefont
		{{Paris-Mandoki}}, \citenamefont {Braun}, \citenamefont {Kumlin},
		\citenamefont {Tresp}, \citenamefont {Mirgorodskiy}, \citenamefont
		{Christaller}, \citenamefont {B{\"u}chler},\ and\ \citenamefont
		{Hofferberth}}]{paris-mandoki_free-space_2017}%
	\BibitemOpen
	\bibfield  {author} {\bibinfo {author} {\bibfnamefont {A.}~\bibnamefont
			{{Paris-Mandoki}}}, \bibinfo {author} {\bibfnamefont {C.}~\bibnamefont
			{Braun}}, \bibinfo {author} {\bibfnamefont {J.}~\bibnamefont {Kumlin}},
		\bibinfo {author} {\bibfnamefont {C.}~\bibnamefont {Tresp}}, \bibinfo
		{author} {\bibfnamefont {I.}~\bibnamefont {Mirgorodskiy}}, \bibinfo {author}
		{\bibfnamefont {F.}~\bibnamefont {Christaller}}, \bibinfo {author}
		{\bibfnamefont {H.~P.}\ \bibnamefont {B{\"u}chler}},\ and\ \bibinfo {author}
		{\bibfnamefont {S.}~\bibnamefont {Hofferberth}},\ }\bibfield  {title}
	{\bibinfo {title} {Free-{{Space Quantum Electrodynamics}} with a {{Single
					Rydberg Superatom}}},\ }\href {https://doi.org/10.1103/PhysRevX.7.041010}
	{\bibfield  {journal} {\bibinfo  {journal} {Physical Review X}\ }\textbf
		{\bibinfo {volume} {7}},\ \bibinfo {pages} {041010} (\bibinfo {year}
		{2017})}\BibitemShut {NoStop}%
	\bibitem [{\citenamefont {Stolz}\ \emph {et~al.}(2022)\citenamefont {Stolz},
		\citenamefont {Hegels}, \citenamefont {Winter}, \citenamefont {R\"ohr},
		\citenamefont {Hsiao}, \citenamefont {Husel}, \citenamefont {Rempe},\ and\
		\citenamefont {D\"urr}}]{Stolz2022}%
	\BibitemOpen
	\bibfield  {author} {\bibinfo {author} {\bibfnamefont {T.}~\bibnamefont
			{Stolz}}, \bibinfo {author} {\bibfnamefont {H.}~\bibnamefont {Hegels}},
		\bibinfo {author} {\bibfnamefont {M.}~\bibnamefont {Winter}}, \bibinfo
		{author} {\bibfnamefont {B.}~\bibnamefont {R\"ohr}}, \bibinfo {author}
		{\bibfnamefont {Y.-F.}\ \bibnamefont {Hsiao}}, \bibinfo {author}
		{\bibfnamefont {L.}~\bibnamefont {Husel}}, \bibinfo {author} {\bibfnamefont
			{G.}~\bibnamefont {Rempe}},\ and\ \bibinfo {author} {\bibfnamefont
			{S.}~\bibnamefont {D\"urr}},\ }\bibfield  {title} {\bibinfo {title}
		{Quantum-logic gate between two optical photons with an average efficiency
			above 40\%},\ }\href {https://doi.org/10.1103/PhysRevX.12.021035} {\bibfield
		{journal} {\bibinfo  {journal} {Phys. Rev. X}\ }\textbf {\bibinfo {volume}
			{12}},\ \bibinfo {pages} {021035} (\bibinfo {year} {2022})}\BibitemShut
	{NoStop}%
	\bibitem [{\citenamefont {Vaneecloo}\ \emph {et~al.}(2022)\citenamefont
		{Vaneecloo}, \citenamefont {Garcia},\ and\ \citenamefont
		{Ourjoumtsev}}]{Vaneecloo2022}%
	\BibitemOpen
	\bibfield  {author} {\bibinfo {author} {\bibfnamefont {J.}~\bibnamefont
			{Vaneecloo}}, \bibinfo {author} {\bibfnamefont {S.}~\bibnamefont {Garcia}},\
		and\ \bibinfo {author} {\bibfnamefont {A.}~\bibnamefont {Ourjoumtsev}},\
	}\bibfield  {title} {\bibinfo {title} {Intracavity rydberg superatom for
			optical quantum engineering: Coherent control, single-shot detection, and
			optical $\ensuremath{\pi}$ phase shift},\ }\href
	{https://doi.org/10.1103/PhysRevX.12.021034} {\bibfield  {journal} {\bibinfo
			{journal} {Phys. Rev. X}\ }\textbf {\bibinfo {volume} {12}},\ \bibinfo
		{pages} {021034} (\bibinfo {year} {2022})}\BibitemShut {NoStop}%
	\bibitem [{\citenamefont {Limonov}\ \emph {et~al.}(2017)\citenamefont
		{Limonov}, \citenamefont {Rybin}, \citenamefont {Poddubny},\ and\
		\citenamefont {Kivshar}}]{limonov_fano_2017}%
	\BibitemOpen
	\bibfield  {author} {\bibinfo {author} {\bibfnamefont {M.~F.}\ \bibnamefont
			{Limonov}}, \bibinfo {author} {\bibfnamefont {M.~V.}\ \bibnamefont {Rybin}},
		\bibinfo {author} {\bibfnamefont {A.~N.}\ \bibnamefont {Poddubny}},\ and\
		\bibinfo {author} {\bibfnamefont {Y.~S.}\ \bibnamefont {Kivshar}},\
	}\bibfield  {title} {\bibinfo {title} {Fano resonances in photonics},\ }\href
	{https://doi.org/10.1038/nphoton.2017.142} {\bibfield  {journal} {\bibinfo
			{journal} {Nature Photonics}\ }\textbf {\bibinfo {volume} {11}},\ \bibinfo
		{pages} {543} (\bibinfo {year} {2017})}\BibitemShut {NoStop}%
	\bibitem [{\citenamefont {Witthaut}\ \emph {et~al.}(2012)\citenamefont
		{Witthaut}, \citenamefont {Lukin},\ and\ \citenamefont
		{S{\o}rensen}}]{witthaut_photon_2012}%
	\BibitemOpen
	\bibfield  {author} {\bibinfo {author} {\bibfnamefont {D.}~\bibnamefont
			{Witthaut}}, \bibinfo {author} {\bibfnamefont {M.~D.}\ \bibnamefont
			{Lukin}},\ and\ \bibinfo {author} {\bibfnamefont {A.~S.}\ \bibnamefont
			{S{\o}rensen}},\ }\bibfield  {title} {\bibinfo {title} {Photon sorters and
			{{QND}} detectors using single photon emitters},\ }\href
	{https://doi.org/10.1209/0295-5075/97/50007} {\bibfield  {journal} {\bibinfo
			{journal} {EPL (Europhysics Letters)}\ }\textbf {\bibinfo {volume} {97}},\
		\bibinfo {pages} {50007} (\bibinfo {year} {2012})}\BibitemShut {NoStop}%
	\bibitem [{\citenamefont {Ralph}\ \emph {et~al.}(2015)\citenamefont {Ralph},
		\citenamefont {S{\"o}llner}, \citenamefont {Mahmoodian}, \citenamefont
		{White},\ and\ \citenamefont {Lodahl}}]{ralph_photon_2015}%
	\BibitemOpen
	\bibfield  {author} {\bibinfo {author} {\bibfnamefont {T.~C.}\ \bibnamefont
			{Ralph}}, \bibinfo {author} {\bibfnamefont {I.}~\bibnamefont {S{\"o}llner}},
		\bibinfo {author} {\bibfnamefont {S.}~\bibnamefont {Mahmoodian}}, \bibinfo
		{author} {\bibfnamefont {A.~G.}\ \bibnamefont {White}},\ and\ \bibinfo
		{author} {\bibfnamefont {P.}~\bibnamefont {Lodahl}},\ }\bibfield  {title}
	{\bibinfo {title} {Photon {{Sorting}}, {{Efficient Bell Measurements}}, and a
			{{Deterministic Controlled- Z Gate Using}} a {{Passive Two-Level
					Nonlinearity}}},\ }\href {https://doi.org/10.1103/PhysRevLett.114.173603}
	{\bibfield  {journal} {\bibinfo  {journal} {Physical Review Letters}\
		}\textbf {\bibinfo {volume} {114}},\ \bibinfo {pages} {173603} (\bibinfo
		{year} {2015})}\BibitemShut {NoStop}%
	\bibitem [{\citenamefont {Yang}\ \emph {et~al.}(2022)\citenamefont {Yang},
		\citenamefont {Lund}, \citenamefont {Pohl}, \citenamefont {Lodahl},\ and\
		\citenamefont {M\o{}lmer}}]{Yang2022}%
	\BibitemOpen
	\bibfield  {author} {\bibinfo {author} {\bibfnamefont {F.}~\bibnamefont
			{Yang}}, \bibinfo {author} {\bibfnamefont {M.~M.}\ \bibnamefont {Lund}},
		\bibinfo {author} {\bibfnamefont {T.}~\bibnamefont {Pohl}}, \bibinfo {author}
		{\bibfnamefont {P.}~\bibnamefont {Lodahl}},\ and\ \bibinfo {author}
		{\bibfnamefont {K.}~\bibnamefont {M\o{}lmer}},\ }\bibfield  {title} {\bibinfo
		{title} {Deterministic photon sorting in waveguide qed systems},\ }\href
	{https://doi.org/10.1103/PhysRevLett.128.213603} {\bibfield  {journal}
		{\bibinfo  {journal} {Phys. Rev. Lett.}\ }\textbf {\bibinfo {volume} {128}},\
		\bibinfo {pages} {213603} (\bibinfo {year} {2022})}\BibitemShut {NoStop}%
	\bibitem [{\citenamefont {Mahmoodian}\ \emph {et~al.}(2018)\citenamefont
		{Mahmoodian}, \citenamefont {{\v C}epulkovskis}, \citenamefont {Das},
		\citenamefont {Lodahl}, \citenamefont {Hammerer},\ and\ \citenamefont
		{S{\o}rensen}}]{mahmoodian_strongly_2018}%
	\BibitemOpen
	\bibfield  {author} {\bibinfo {author} {\bibfnamefont {S.}~\bibnamefont
			{Mahmoodian}}, \bibinfo {author} {\bibfnamefont {M.}~\bibnamefont {{\v
					C}epulkovskis}}, \bibinfo {author} {\bibfnamefont {S.}~\bibnamefont {Das}},
		\bibinfo {author} {\bibfnamefont {P.}~\bibnamefont {Lodahl}}, \bibinfo
		{author} {\bibfnamefont {K.}~\bibnamefont {Hammerer}},\ and\ \bibinfo
		{author} {\bibfnamefont {A.~S.}\ \bibnamefont {S{\o}rensen}},\ }\bibfield
	{title} {\bibinfo {title} {Strongly {{Correlated Photon Transport}} in
			{{Waveguide Quantum Electrodynamics}} with {{Weakly Coupled Emitters}}},\
	}\href {https://doi.org/10.1103/PhysRevLett.121.143601} {\bibfield  {journal}
		{\bibinfo  {journal} {Physical Review Letters}\ }\textbf {\bibinfo {volume}
			{121}},\ \bibinfo {pages} {143601} (\bibinfo {year} {2018})}\BibitemShut
	{NoStop}%
	\bibitem [{\citenamefont {Prasad}\ \emph {et~al.}(2020)\citenamefont {Prasad},
		\citenamefont {Hinney}, \citenamefont {Mahmoodian}, \citenamefont {Hammerer},
		\citenamefont {Rind}, \citenamefont {Schneeweiss}, \citenamefont
		{S{\o}rensen}, \citenamefont {Volz},\ and\ \citenamefont
		{Rauschenbeutel}}]{prasad_correlating_2020}%
	\BibitemOpen
	\bibfield  {author} {\bibinfo {author} {\bibfnamefont {A.~S.}\ \bibnamefont
			{Prasad}}, \bibinfo {author} {\bibfnamefont {J.}~\bibnamefont {Hinney}},
		\bibinfo {author} {\bibfnamefont {S.}~\bibnamefont {Mahmoodian}}, \bibinfo
		{author} {\bibfnamefont {K.}~\bibnamefont {Hammerer}}, \bibinfo {author}
		{\bibfnamefont {S.}~\bibnamefont {Rind}}, \bibinfo {author} {\bibfnamefont
			{P.}~\bibnamefont {Schneeweiss}}, \bibinfo {author} {\bibfnamefont {A.~S.}\
			\bibnamefont {S{\o}rensen}}, \bibinfo {author} {\bibfnamefont
			{J.}~\bibnamefont {Volz}},\ and\ \bibinfo {author} {\bibfnamefont
			{A.}~\bibnamefont {Rauschenbeutel}},\ }\bibfield  {title} {\bibinfo {title}
		{Correlating photons using the collective nonlinear response of atoms weakly
			coupled to an optical mode},\ }\href
	{https://doi.org/10.1038/s41566-020-0692-z} {\bibfield  {journal} {\bibinfo
			{journal} {Nature Photonics}\ }\textbf {\bibinfo {volume} {14}},\ \bibinfo
		{pages} {719} (\bibinfo {year} {2020})}\BibitemShut {NoStop}%
	\bibitem [{\citenamefont {Mahmoodian}\ \emph {et~al.}(2020)\citenamefont
		{Mahmoodian}, \citenamefont {Calaj{\'o}}, \citenamefont {Chang},
		\citenamefont {Hammerer},\ and\ \citenamefont
		{S{\o}rensen}}]{mahmoodian_dynamics_2020}%
	\BibitemOpen
	\bibfield  {author} {\bibinfo {author} {\bibfnamefont {S.}~\bibnamefont
			{Mahmoodian}}, \bibinfo {author} {\bibfnamefont {G.}~\bibnamefont
			{Calaj{\'o}}}, \bibinfo {author} {\bibfnamefont {D.~E.}\ \bibnamefont
			{Chang}}, \bibinfo {author} {\bibfnamefont {K.}~\bibnamefont {Hammerer}},\
		and\ \bibinfo {author} {\bibfnamefont {A.~S.}\ \bibnamefont {S{\o}rensen}},\
	}\bibfield  {title} {\bibinfo {title} {Dynamics of {{Many-Body Photon Bound
					States}} in {{Chiral Waveguide QED}}},\ }\href
	{https://doi.org/10.1103/PhysRevX.10.031011} {\bibfield  {journal} {\bibinfo
			{journal} {Physical Review X}\ }\textbf {\bibinfo {volume} {10}},\ \bibinfo
		{pages} {031011} (\bibinfo {year} {2020})}\BibitemShut {NoStop}%
	\bibitem [{\citenamefont {Iversen}\ and\ \citenamefont
		{Pohl}(2021)}]{iversen_strongly_2021}%
	\BibitemOpen
	\bibfield  {author} {\bibinfo {author} {\bibfnamefont {O.~A.}\ \bibnamefont
			{Iversen}}\ and\ \bibinfo {author} {\bibfnamefont {T.}~\bibnamefont {Pohl}},\
	}\bibfield  {title} {\bibinfo {title} {Strongly {{Correlated States}} of
			{{Light}} and {{Repulsive Photons}} in {{Chiral Chains}} of {{Three-Level
					Quantum Emitters}}},\ }\href {https://doi.org/10.1103/PhysRevLett.126.083605}
	{\bibfield  {journal} {\bibinfo  {journal} {Physical Review Letters}\
		}\textbf {\bibinfo {volume} {126}},\ \bibinfo {pages} {083605} (\bibinfo
		{year} {2021})}\BibitemShut {NoStop}%
	\bibitem [{\citenamefont {Iversen}\ and\ \citenamefont
		{Pohl}(2022)}]{iversen_self-ordering_2022}%
	\BibitemOpen
	\bibfield  {author} {\bibinfo {author} {\bibfnamefont {O.~A.}\ \bibnamefont
			{Iversen}}\ and\ \bibinfo {author} {\bibfnamefont {T.}~\bibnamefont {Pohl}},\
	}\bibfield  {title} {\bibinfo {title} {Self-ordering of individual photons in
			waveguide {{QED}} and {{Rydberg-atom}} arrays},\ }\href
	{https://doi.org/10.1103/PhysRevResearch.4.023002} {\bibfield  {journal}
		{\bibinfo  {journal} {Physical Review Research}\ }\textbf {\bibinfo {volume}
			{4}},\ \bibinfo {pages} {023002} (\bibinfo {year} {2022})}\BibitemShut
	{NoStop}%
\end{thebibliography}
\end{document}


\begin{CJK*}{UTF8}{gbsn}
	\title{Supplemental material to Quantum nonlinear optics in atomic dual arrays} 
	\author{Simon Panyella Pedersen}
	\email{spp@phys.au.dk}
	\affiliation{\affA}
	\author{Lida Zhang~(\CJKfamily{gbsn}张理达)}
	\email{zhanglida@phys.au.dk}
	\affiliation{\affA}
	\author{Thomas Pohl}
	\email{pohl@phys.au.dk}
	\affiliation{\affA}

\maketitle
\end{CJK*}
\section{Transmission resonances of dual atomic arrays}
The Greens function tensor ${\bf G}({\bf R})$ of the free-space electromagnetic field has cartesian elements given by
\begin{align}
	G_{ij}({\bf R}) = \frac{e^{ikR}}{4\pi R}\left[\left(1 + \frac{ikR - 1}{k^2R^2}\right)\delta_{ij} - \left(1 + 3\frac{ikR - 1}{k^2R^2}\right)\frac{R_{i}R_{j}}{R^2}\right] \label{eq:Gcomponents},
\end{align}
where $R=|{\bf R}|$ and $R_{i}$ ($i=1,2,3$) denotes the cartesian component of the three-dimensional distance vector between a given point in space and a radiating dipole. The Greens function yields the coefficients $J_{nm}$ and $\Gamma_{nm}$ in Eq. (1) of the main text via $ J_{nm} + i\Gamma_{nm} = \mathcal{G}_{nm} $, where
\begin{align}
	\mathcal{G}_{nm} = \frac{6\pi\gamma}{k}{\bf e}^{\dagger}_{+}\cdot{\bf G}({\bf R}_{n} - {\bf R}_{m})\cdot{\bf e}_{+}, \label{eq:Gnm}
\end{align}
and ${\bf R}_{n}$ and ${\bf R}_{m}$ denote the three-dimensional positions of two atoms in the arrays. Considering two parallel arrays, positioned in the $x-y$ plane, we use the notation ${\bf R}_n=(x_n,y_n,z_n)=({\bf r}_n,z_n)$. As discussed in the main text, we focus on a single atomic transition driven by circularly polarized light, choosing $ {\bf e}_{+} = (1,i,0)^{T}/\sqrt{2} $ as a specific polarization vector. Explicitly, one, therefore, obtains 
\begin{align} 
	\mathcal{G}_{nm}(\ell) = \frac{3\gamma}{2}\frac{e^{ikR_{nm}}}{k R_{nm}}\left[1 + \frac{ikR_{nm} - 1}{k^2R_{nm}^2} - \left(1 + 3\frac{ikR_{nm} - 1}{k^2R_{nm}^2}\right)\frac{r_{nm}^2}{2R_{nm}^2}\right] \label{eq:GL},
\end{align}
where $R_{nm}^2=r_{nm}^2+\ell^2$, and $\ell$ denotes the atomic separation along the $z$-axis. Hence, $\ell=0$ describes interactions between two atoms in the same array and $\ell>0$ corresponds to interactions between atoms in each of the two arrays. We employ \cref{eq:GL} together with Eq. (1) of the main text for the numerical simulations of finite lattices with planar and non-planar geometries.

For the analytical estimates, discussed in the main text, we consider two parallel infinitely extended arrays, positioned at $z=\pm L$, respectively.
The Fourier transform of \cref{eq:GL} in the $x-y$ plane 
\begin{align}
	\begin{split}
		\tilde{\mathcal{G}}_{{\bf k}_{\perp}}(\ell) &\equiv -\tilde{\Delta}_\ell+i\tilde{\Gamma}_\ell=\sum_{n}\mathcal{G}_{n0}(\ell)e^{-i{\bf k}_{\perp}\cdot{\bf r}_{n}}\\
		&= \frac{6\pi\gamma}{k}\left(\frac{i}{2k^2a^2}\sum_{{\bf q}}\frac{k^2 - ({\bf k}_{\perp} - {\bf q})^2/2}{\sqrt{k^2 - ({\bf k}_{\perp} - {\bf q})^2}}e^{i\sqrt{k^2 - ({\bf k}_{\perp} - {\bf q})^2}\ell} - \delta_{\ell,0}\Re[{\bf e}_{+}^{*}\cdot{\bf G}(0)\cdot{\bf e}_{+}]\right) , \label{eq:FTG_L}
	\end{split}
\end{align}
then defines contributions to the effective level shifts and decay rates from interactions between atoms in the same array ($\ell=0$) and in different arrays ($\ell=L$). Here, the sum runs over the reciprocal lattice sites, $ {\bf q} = (2\pi/a){\bf m} $ with $ {\bf m} = (m_{x},m_{y}) $ and $ m_{x},m_{y} \in \mathbb{Z} $. One can see from this expression that $ a < \lambda/2 $ ensures that the imaginary part of $ \tilde{\mathcal{G}}$ has only one contribution  from the ${\bf q} = 0 $ term, since $ |{\bf k}_{\perp}| < k $. This implies that the transverse momentum of the incident light is conserved by the interaction with the array, eliminating any losses from the incident mode due to photon scattering. If ${\bf k}_{\perp}=0$, this condition is already met for $a<\lambda$. Divergences in the above expression originate from the single-atom Lamb shift and can be accounted for the in the detuning $\Delta$. The expression can be evaluated numerically using a regularization procedure described in \cite{shahmoon_cooperative_2017}.

For plane wave driving at normal incidence (${\bf k}_\perp=0$), one obtains the simple result
\begin{subequations}
	\begin{align}
		\tilde{\Delta}_{\ell} &= \tilde{\Gamma}\left(\sin(k\ell) + \frac{1}{k}\sum_{{\bf q} \neq 0}\frac{|{\bf q}|^2/2 - k^2}{\sqrt{|{\bf q}|^2 - k^2}}e^{-\sqrt{|{\bf q}|^2 - k^2}\ell}\right) \label{eq:FDeltaL},\\
		\tilde{\Gamma}_{\ell} &= \tilde{\Gamma}\cos(k\ell) . \label{eq:FGammaL}
	\end{align}
\end{subequations}
The intra-layer terms are given by
\begin{align}
	\begin{split}\label{eq:expliticJGamma}
		\tilde{\Gamma}_0 &\equiv\tilde{\Gamma} = \frac{3\pi\gamma}{k^2a^2}, \\
				\tilde{\Delta}_0 &\equiv\tilde{\Delta} = \frac{\tilde{\Gamma}}{k}\sum_{{\bf m} \neq 0}\frac{(2\pi{\bf m}/a)^2/2 - k^2}{\sqrt{(2\pi{\bf m}/a)^2 - k^2}} - \frac{6\pi\gamma}{k}\Re\left[{\bf e}^{\dagger}_{+}\cdot{\bf G}(0)\cdot{\bf e}_{+}\right] .
	\end{split}
\end{align} 
The inter-layer contribution, $\tilde{\Delta}_{L}$, to the collective shift consists of two terms. The first describes the energy shift due to exchange of photons that propagate between the two arrays and therefore oscillates at the wavelength, $\lambda=2\pi/k$, of the transition resonance. The second term, decays rapidly with $L$ and describes the direct near-field interaction between the two arrays. Either term dominates in the respective limits of large and small array distances, L, as discussed in the main text and as we shall describe in more detail below. 

To this end, let us denote the atomic transition operators for the first and second array with respect to the field incidence by $ \hat{\sigma}_{n}^{(1)} $ and $ \hat{\sigma}_{n}^{(2)} $, respectively. One can then introduce (anti)symmetric superpositions $\hat{\sigma}_{\pm,n}=(\hat{\sigma}_{n}^{(1)}\pm\hat{\sigma}_{n}^{(2)})/\sqrt{2}$ at a given lattice site ${\bf r}_n$ of the dual array, which define the symmetric and antisymmetric dimer states $|\pm\rangle_n=\hat{\sigma}_{\pm,n}^\dagger|0\rangle$, discussed in the main text. For plane wave driving with ${\bf k}_\perp=0$ we can write the Rabi frequencies for each array as $\Omega_n^{(1)}=\Omega e^{-ikL/2}$ and $\Omega_n^{(2)}=\Omega e^{ikL/2}$, with a real and constant amplitude $\Omega$. This yields 
\begin{subequations}
	\begin{align}
	\begin{split}
		\hat{H} =& -\sum_{n}[\Delta + J_{nn}(L)]\hat{\sigma}_{+,n}^{\dagger}\hat{\sigma}_{+,n} - \sum_{n}(\sqrt{2}\Omega \cos(kL/2)\hat{\sigma}_{+,n} + {\rm h.c.}) - \sum_{n \neq m}[J_{nm}(0) + J_{nm}(L)]\hat{\sigma}_{+,n}^{\dagger}\hat{\sigma}_{+,m}\\
		& -\sum_{n}[\Delta - J_{nn}(L)]\hat{\sigma}_{-,n}^{\dagger}\hat{\sigma}_{-,n} - i\sum_{n}(\sqrt{2}\Omega \sin(kL/2)\hat{\sigma}_{-,n} + {\rm h.c.}) - \sum_{n \neq m}[J_{nm}(0) - J_{nm}(L)]\hat{\sigma}_{-,n}^{\dagger}\hat{\sigma}_{-,m}, \label{eq:Hpm}
	\end{split}\\
	\begin{split}
		\mathcal{L}[\hat{\rho}] =& \sum_{n,m}[\Gamma_{nm}(0) + \Gamma_{nm}(L)]\left(2\hat{\sigma}_{+,n}^{\dagger}\hat{\rho}\hat{\sigma}_{+,m} - \{\hat{\sigma}_{+,n}^{\dagger}\hat{\sigma}_{+,m},\hat{\rho}\}\right)\\
		& + \sum_{n,m}[\Gamma_{nm}(0) - \Gamma_{nm}(L)]\left(2\hat{\sigma}_{-,n}^{\dagger}\hat{\rho}\hat{\sigma}_{-,m} - \{\hat{\sigma}_{-,n}^{\dagger}\hat{\sigma}_{-,m},\hat{\rho}\}\right), \label{eq:Lpm}
	\end{split}
\end{align}
\end{subequations}
where the interaction coefficients are obtained from $J_{nm}(\ell) + i\Gamma_{nm}(\ell) = \mathcal{G}_{nm}(\ell)$ and \cref{eq:GL}. The parity symmetry of the light-induced interactions implies that there is no direct coupling between symmetric and antisymmetric states, such that only the external driving field can transfer populations between the two parity eigenstates at a given lattice site. 
The Hamiltonian can be rewritten as 
\begin{subequations}
	\begin{align}
		\hat{H} &= -\left(\Delta - \tilde{\Delta}_{+}\right)\tilde{\sigma}_+^{\dagger}\tilde{\sigma}_+ - \left(\Delta - \tilde{\Delta}_{-}\right)\tilde{\sigma}_-^{\dagger}\tilde{\sigma}_- - \left(\tilde{\Omega}_{+}\tilde{\sigma}_+ + i\tilde{\Omega}_{-}\tilde{\sigma}_- + {\rm h.c.}\right) , \label{eq:diagH_dual}\\
		\mathcal{L}[\hat{\rho}] &= \tilde{\Gamma}_{+}\left(2\tilde{\sigma}_+\hat{\rho}\tilde{\sigma}_+^{\dagger} - \left\{\tilde{\sigma}_+^{\dagger}\tilde{\sigma}_+,\hat{\rho}\right\}\right) + \tilde{\Gamma}_{-}\left(2\tilde{\sigma}_-^{\dagger}\hat{\rho}\tilde{\sigma}_- - \left\{\tilde{\sigma}_-^{\dagger}\tilde{\sigma}_-,\hat{\rho}\right\}\right), \label{eq:diagL_dual}
	\end{align}
\end{subequations}
where $\tilde{\Omega}_{+} = \sqrt{2}\tilde{\Omega}\cos(kL/2)$, $\tilde{\Omega}_{-} = \sqrt{2}\tilde{\Omega}\sin(kL/2)$, $\tilde{\Omega}=\sqrt{N}\Omega$, $ \tilde{\Delta}_{\pm} = \tilde{\Delta} \pm \tilde{\Delta}_{L} $, and $ \tilde{\Gamma}_{\pm} = \tilde{\Gamma} \pm \tilde{\Gamma}_{L}= \tilde{\Gamma}[1\pm\cos(kL)]$, as given in the main text. The operators $\tilde{\sigma}_\pm=N^{-1/2}\sum_n \hat{\sigma}_{\pm,n}$ generate the collective (anti)symmetric states 
\begin{equation}
|\pm\rangle=\tilde{\sigma}_\pm^\dagger|0\rangle=N^{-1/2}\sum_n|\pm\rangle_n,
\end{equation}
discussed in the main text. Using a truncated basis with not more than a single excitation, the system can be described by a simple non-Hermitian Hamiltonian
\begin{align}
	\begin{split}
		\hat{H} =& -\left(\Delta - \tilde{\Delta}_{+}+i\tilde{\Gamma}_+\right)|+\rangle\langle+| - \left(\Delta - \tilde{\Delta}_{-}+i\tilde{\Gamma}_-\right)|-\rangle\langle-| \\
		& - \left(\sqrt{2}\tilde{\Omega}\cos(kL/2) |0\rangle\langle+| + i\sqrt{2}\tilde{\Omega}\sin(kL/2) |0\rangle\langle-| + {\rm h.c.}\right). \label{eq:Heff}
	\end{split}
\end{align}
From the corresponding steady state of the driven atoms, one can determine the transmitted field from Eq. (2) of the main text as
\begin{align}
	\begin{split}
		\hat{E} &=E_{\rm in}+i\frac{a^2\tilde{\Gamma}}{\eta d}\sum_n (e^{ikL/2}\hat{\sigma}_n^{(1)}+e^{-ikL/2}\hat{\sigma}_n^{(2)})\\
		&= E_{\rm in}+E_{\rm in}\frac{i\tilde{\Gamma}}{\tilde{\Omega}}\sqrt{2}(\cos(kL/2)\tilde{\sigma}_{+}+i\sin(kL/2)\tilde{\sigma}_{-})
	\end{split}
\end{align}
to obtain the transmission amplitude of the dual array
\begin{equation}\label{eq:TL}
T=1-i\left(\frac{\tilde{\Gamma}_+}{\Delta-\tilde{\Delta}_++i\tilde{\Gamma}_+}+\frac{\tilde{\Gamma}_-}{\Delta-\tilde{\Delta}_-+i\tilde{\Gamma}_-}\right).
\end{equation}
The transmission obtained from this expression is shown in Fig. 1a of the main text. Below we discuss the limiting behaviour of this simple expression for small and large values of $L$, respectively.

\subsection{Small-$L$ limit}
\begin{figure}
	\centering
	\includegraphics[width=0.8\textwidth]{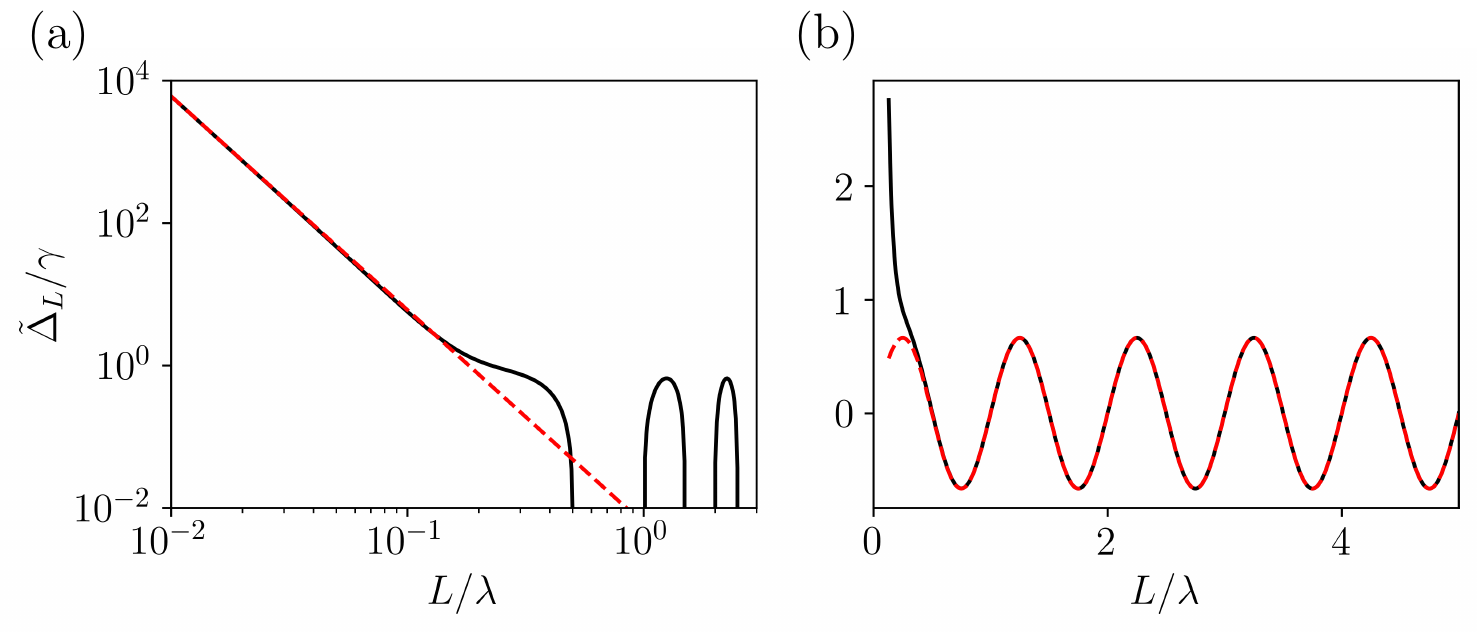}
	\caption{\label{fig1}  The collective energy shift $\tilde{\Delta}_{L} $ for two arrays with a lattice spacing of $ a = 0.6\lambda $ as a function of their distance $ L $. Panel (a) demonstrates good agreement between the numerical evaluation of \cref{eq:FDeltaL} (black solid line) and the dipolar scaling $\tilde{\Delta}_L\approx3\gamma/2(kL)^3$ (red dashed line) for small $ L $.  Panel (b) compares the exact numerical result (black solid line) to the large-distance limit  $\tilde{\Delta}_L\approx\tilde{\Gamma}\sin(kL)$ (red dashed line).}
\end{figure}

For small values of $L<a$, the major contribution to the interaction stems from atomic pairs at identical sites in different arrays. We thus obtain from \cref{eq:GL},
\begin{align}
	G_{nn}(L) = \frac{3\gamma}{2}\frac{e^{ikL}}{k L}\left(1 + \frac{ikL - 1}{k^2L^2}\right)\approx  -\frac{3\gamma}{2(k L)^3}\label{eq:GsmallL},
\end{align}
that the energy shift $\tilde{\Delta}_L\approx 3\gamma/2(kL)^3$ predominantly arises from the direct dipole-dipole interaction between adjacent atoms from each array. In \cref{fig1}, we compare this expression to the numerical Fourier transform \cref{eq:FDeltaL} and find good agreement for small $L$. This implies that the two resonances at $\tilde{\Delta}_\pm\approx \tilde{\Delta}\pm 3\gamma/2(kL)^3$ in \cref{eq:TL} are energetically well separated, such that we find from \cref{eq:TL} two individual transmission resonances
\begin{equation}
T=1-i\left(\frac{\tilde{\Gamma}_\pm}{\Delta-\tilde{\Delta}_\pm+i\tilde{\Gamma}_\pm}\right)
\end{equation}
as discussed in the main text and as shown in Fig. 1 of the article. As $L$ decreases the width of the subradiant resonance vanishes as $\tilde{\Gamma}_-\approx\frac{\tilde{\Gamma}}{2}(kL)^2$, while we find a superradiant rate $\tilde{\Gamma}_+\approx2\tilde{\Gamma}$ for the resonance at positive frequency detunings. These approximate results agree well with the numerical findings shown in Fig. 1a of the main text.

\subsection{Large-$L$ limit}
\begin{figure}
	\centering
	\includegraphics[width=0.7\textwidth]{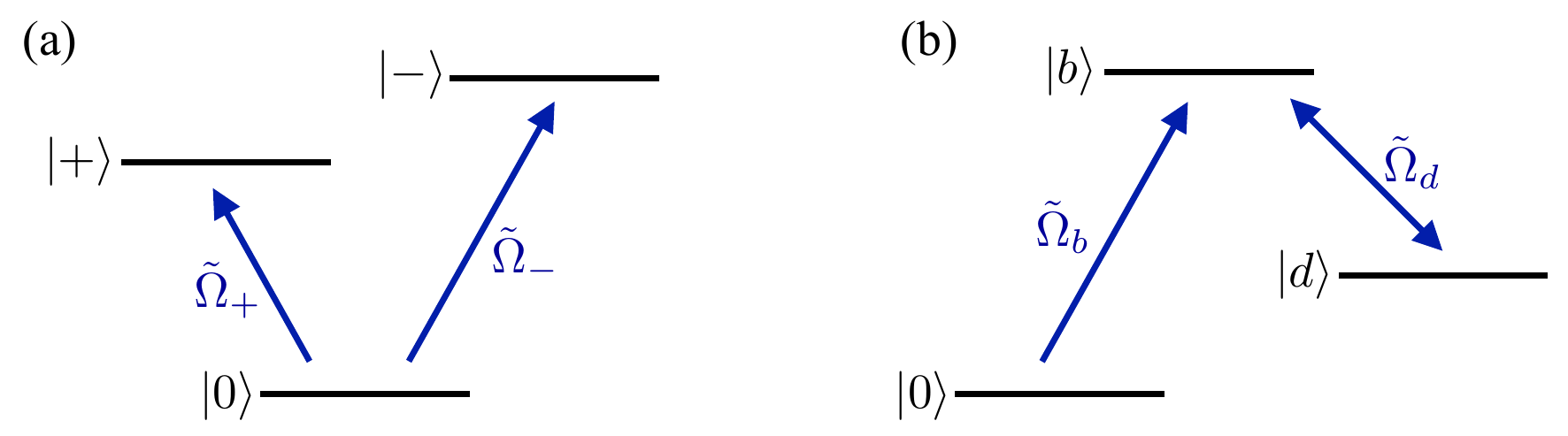}
	\caption{\label{fig2} (a) Effective level scheme according to the Hamiltonian \cref{eq:diagH_dual}  in terms of the non-interacting symmetric and antisymmetric collective states, $|\pm\rangle$, which are both excited from the ground state $|0\rangle$ by the incident light field. (b) Effective lambda-type level scheme according to the Hamiltonian \cref{eq:Hbd} in terms of the collective bright and dark states. Only the bright state $|b\rangle$ is driven by the light field with a Rabi frequency $\tilde{\Omega}_b$, while the coupling, $\tilde{\Omega}_d$ arises from the atomic interaction.}
\end{figure}

For large distances between the arrays, we can neglect their evanescent field coupling, such that $\tilde{\Delta}_L\approx\tilde{\Gamma}\sin(kL)$, according to \cref{eq:FDeltaL}, follows a simple sinusoidal oscillation (cf. \cref{fig1}). Since, the two reflection resonances in \cref{eq:TL} now overlap significantly they both contribute to the linear response of the array. We can gain further insights by transforming to a new basis generated by $ \tilde{B} = \cos(kL/2)\tilde{\sigma}_+ + i\sin(kL/2)\tilde{\sigma}_- $ and $ \tilde{D} = -\sin(kL/2)\tilde{\sigma}_+ + i\cos(kL/2)\tilde{\sigma}_- $. The state
\begin{equation}
|b\rangle=\tilde{B}^\dagger|0\rangle
\end{equation}
corresponds to a bright state with maximum coupling to the incident driving field, while
\begin{equation}
|d\rangle=\tilde{D}^\dagger|0\rangle,
\end{equation}
represents a dark state with respect to the external driving field that is only coupled to $|b\rangle$ via the atomic interactions with a coupling strength $\tilde{\Delta}_{L}\sin(kL)$. From \cref{eq:diagH_dual,eq:diagL_dual} we obtain
\begin{subequations}
	\begin{align}
		\begin{split}
			\hat{H} =& -\left(\Delta - \tilde{\Delta} - \tilde{\Gamma}\sin(kL)\cos(kL)\right)\tilde{B}^{\dagger}\tilde{B} - \left(\Delta - \tilde{\Delta} + \tilde{\Gamma}\sin(kL)\cos(kL)\right)\tilde{D}^{\dagger}\tilde{D}\\
			& + \tilde{\Omega}_d \left(\tilde{B}^{\dagger}\tilde{D} + \tilde{D}^{\dagger}\tilde{B}\right) - \left(\tilde{\Omega}_b\tilde{B} + {\rm h.c.}\right) , \label{eq:Hbd}
		\end{split}\\
		\begin{split}
			\mathcal{L}[\hat{\rho}] =& \left(\tilde{\Gamma} + \tilde{\Gamma}\cos^2(kL)\right)\left(2\tilde{B}\hat{\rho}\tilde{B}^{\dagger} - \left\{\tilde{B}^{\dagger}\tilde{B},\hat{\rho}\right\}\right) + \left(\tilde{\Gamma}\sin^2(kL)\right)\left(2\tilde{D}^{\dagger}\hat{\rho}\tilde{D} - \left\{\tilde{D}^{\dagger}\tilde{D},\hat{\rho}\right\}\right)\\
			& - \tilde{\Gamma}\cos^2(kL)\left(2\tilde{B}^{\dagger}\hat{\rho}\tilde{D} + 2\tilde{D}^{\dagger}\hat{\rho}\tilde{B} - \left\{\tilde{B}^{\dagger}\tilde{D} + \tilde{D}^{\dagger}\tilde{B},\hat{\rho}\right\}\right) ,
		\end{split}
	\end{align}
\end{subequations}
where $\tilde{\Omega}_b=\sqrt{2}\tilde{\Omega}$ and $\tilde{\Omega}_d=\tilde{\Gamma}\sin^2(kL)$. Here, we have used $\tilde{\Gamma}_L=\tilde{\Gamma}\cos(kL)$ and the large-$L$ limit $\tilde{\Delta}_L=\tilde{\Gamma}\sin(kL)$. This describes an effective ladder-type three-level system that is illustrated in \cref{fig2}. While the transmitted field 
\begin{equation}\label{eqS:Eout}
\hat{E}= E_{\rm in}+E_{\rm in}\frac{i\tilde{\Gamma}}{\tilde{\Omega}}\sqrt{2}(\cos(kL/2)\tilde{\sigma}_{+}+i\sin(kL/2)\tilde{\sigma}_{-})= E_{\rm in}+E_{\rm in}\frac{\sqrt{2}i\tilde{\Gamma}}{\tilde{\Omega}}\tilde{B}
\end{equation}
is determined by the bright state only, the interference with the dark-state coupling can nevertheless generate transmission resonances akin to electromagnetically induced transparency in three-level systems \cite{fleischhauer_electromagnetically_2005}. From the steady state 
\begin{equation}\label{eqS:B}
\langle\tilde{B}\rangle=-\frac{\tilde{\Omega}}{\sqrt{2}}\left(\frac{1+\cos(kL)}{\Delta-\tilde{\Delta}-\tilde{\Gamma}\sin(kL)+i\tilde{\Gamma}(1+\cos(kL))}+\frac{1-\cos(kL)}{\Delta-\tilde{\Delta}+\tilde{\Gamma}\sin(kL)+i\tilde{\Gamma}(1-\cos(kL))}\right)
\end{equation}
one finds that the interference between the two excitation pathways leads to a simple expression
\begin{equation}
\langle \tilde{B} \rangle= i\frac{\tilde{\Omega}}{\sqrt{2}\tilde{\Gamma}}(1+e^{-2ikL})
\end{equation}
if 
\begin{equation}\label{eq:condition}
\tilde{\Delta}-\Delta = \tilde{\Gamma}\tan kL,
\end{equation}
which corresponds to Eq. (5) in the main text. Under this condition we have 
\begin{equation}
T=1+i\frac{\sqrt{2}\tilde{\Gamma}}{\tilde{\Omega}}\langle \tilde{B}\rangle= -e^{-2ikL}.
\end{equation}
The interference between the two excitation pathways, thus, gives rise to a transmission resonance, where the medium becomes fully transparent, $|T|^2=1$. While this is indeed similar to electromagnetically induced transparency in three-level systems with a ladder-type coupling scheme, the atomic polarization, $\sim \langle \tilde{B}\rangle $, only vanishes for $kL=\pm n \pi/2$ (for integer values of $n$).
In the present case, we still find obtain perfect transparency for finite values of $\langle \tilde{B}\rangle $ such that the transmitted field acquires a tunable phase of $\pi-2kL$ at no photon losses. Moreover, the width of this transmission resonance depends on $\Delta$ and vanishes upon approaching $\Delta=\tilde{\Delta}$. As we shall discuss below, this gives rise to a long confinement time of incident photons within the formed resonator and consequently strong photon-photon interactions.

The transmission $T=\langle\hat{E}\rangle/E_{\rm in}$ obtained from \cref{eqS:Eout,eqS:B} can be expressed as 
\begin{equation}
		T= \frac{t^2}{1 - r^{2}e^{2ikL}} 
\end{equation}
in terms of the transmission and reflection amplitudes of each individual array, given by Eq. (3) of the main text. Expectedly, this expression coincides with the transmission coefficient of a Fabry-P\'{e}rot resonator composed of two identical mirrors with transmission and reflection amplitude $ t $ and $ r $ placed at a distance $ L $.

\section{Group delay of transmitted photons}
The interaction of the incident light with the atoms in general leads to a group delay.
One can determine the corresponding delay time, from the derivative of the transmission amplitude with respect to the frequency of the field according to \cite{facchinetti_storing_2016}
\begin{align}
	\tau = {\rm Im}\left(\frac{1}{T}\frac{{\rm d}T}{{\rm d}\Delta}\right).
\end{align}
For a single array, this yields
\begin{align}
	\tau = \frac{\tilde{\Gamma}}{(\Delta - \tilde{\Delta})^2 + \tilde{\Gamma}^2} = \frac{|r|^2}{\tilde{\Gamma}}.
\end{align}
Expectedly, the delay time is simply given by the inverse decay rate on the reflection resonance. Equivalently, we can calculate the delay time for the dual-array setting. While the general expression is rather lengthy, one can find a simple form in the two limiting cases of small and large values of $L$. For small $L$ we obtain
\begin{align}
	\tau_\pm \approx \frac{|R|^2}{\tilde{\Gamma}_{\pm}}
\end{align}
on the two reflection resonances at $\Delta=\Delta_\pm$, which is consistent with the interpretation in terms of an effective single array composed of either subradiant or superradiant atomic dimers, as discussed in the main text. Here, $ |R|^2 = 1-|T|^2 $ is the reflection coefficient of the dual array. For large array distances, where we can neglect the evanescent-field coupling between the arrays we obtain instead
\begin{align}
	\tau = \frac{2}{\Delta-\tilde{\Delta}}{\rm Im}\left(\frac{i\delta - 1 + e^{2ikL}}{(\delta + i)^2 + e^{2ikL}}\right)
\end{align}
where $ \delta = (\Delta - \tilde{\Delta})/\tilde{\Gamma} $. Along the transmission maximum defined by \cref{eq:condition}, this yields the delay time given in Eq. (6) of the main text.

\section{Finite size effects}
In this section we provide additional details on the behaviour of finite arrays that will be relevant for future experiments and which have been used in our numerical simulations of the nonlinear optical response.

\subsection{Gaussian driving beam}
\begin{figure}
	\centering
	\includegraphics[width=0.7\textwidth]{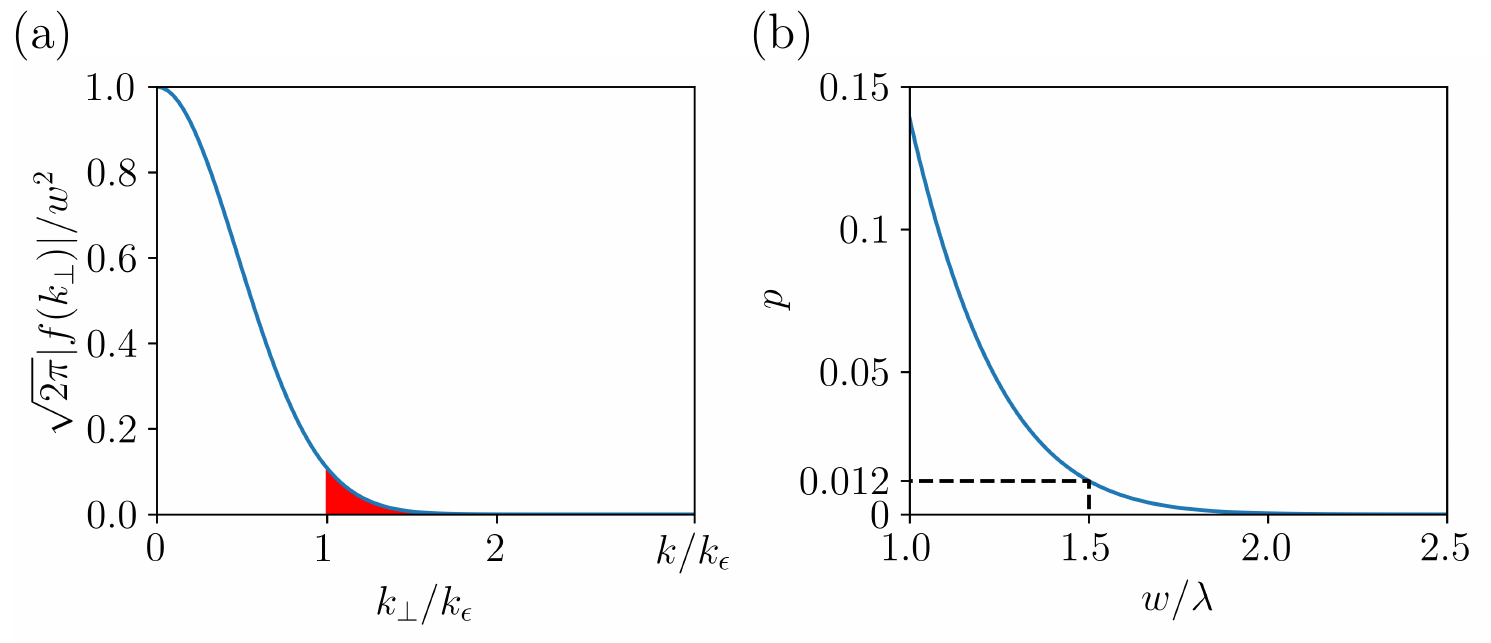}
	\caption{\label{fig3} (a) Gaussian mode function $\sqrt{2\pi}|f(k_{\perp},0)|/w^2$ for a beam waist $w=1.5\lambda$. The momentum is scaled by $k_\epsilon=\sqrt{2\epsilon}k$ with $\epsilon=0.05$, as described in the text. (b) Relative fraction $p=\int_{k_{\epsilon}}^\infty{\rm d}k_\perp k_\perp |f(k_\perp,0)|^2/2\pi w^2$ as a function of $w$ for $\epsilon=0.05$.}
\end{figure}

For our numerical simulations we choose the following normalized Gaussian mode for the in-coming field and the detection mode
\begin{align}
	f({\bf r}) = \sqrt{\frac{2}{\pi}}\frac{w}{w(z)}\exp\left[\frac{-(x^2 + y^2)}{w(z)^2}\right]\exp\left[i\left(kz + k\frac{x^2 + y^2}{2R(z)} - \psi(z)\right)\right] ,
\end{align}
where $ w(z) = w\sqrt{1 + \left(\frac{z}{z_{R}}\right)^2} $ is the beam waist at $ z $, $ z_{R} = \pi w^2/\lambda $ is the Rayleigh range, $ R(z) = z\left[1 + \left(z_{R}/z\right)^2\right] $ is the $z$-dependent radius of the wavefront curvature of the beam. Moreover, $ \psi(z) = \arctan\left(z/z_{R}\right) $ denotes the Gouy phase. In $ (k_{x},k_{y},z) $-space (i.e. Fourier transformed with respect to $ x$ and $y $ only) the mode takes on a more simple form
\begin{align}\label{eq:ftilde}
	\tilde{f}({\bf k}_\perp,z) = \tilde{g}({\bf k}_\perp)e^{ik_z z}=\sqrt{2\pi}w^2e^{-k_\perp^2w^2/4}e^{ik\left(1 - \frac{k_\perp^2}{2k^2}\right)z} ,
\end{align}
where ${\bf k}_\perp=(k_x,k_y)$, $ k_{\perp}^2 = k_{x}^2 + k_{y}^2 $, and $ k_{z} = \sqrt{k^2 - k_{\perp}^2} \approx k\left[1 - k_{\perp}^{2}/(2k^2)\right] $. 
This form of the field mode satisfies the paraxial wave equation. The paraxial approximation is valid if $ k_{z} \approx k\left[1 - k_{\perp}^{2}/2k^2\right] $, i.e. if $k_\perp^2 /2k^2=\epsilon\ll 1$. In our simulations we use $w=1.5\lambda$, for which this approximation is well justified, as illustrated in \cref{fig3}. In fact, for $\epsilon=0.05$, the mode function is already suppressed by an order of magnitude and the relative fraction $p=\int_{k_{\epsilon}}^\infty{\rm d}k_\perp k_\perp |f(k_\perp,0)|^2/2\pi w^2=e^{-k_{\epsilon}^2w^2/2}$ of momenta that would give $\epsilon>0.05$ is $p\sim 0.01$. Hence, the paraxial approximation gives reliable results for $ w \sim 1.5\lambda $ and further improves rapidly for larger values of the beam waist.

\subsection{Diffraction losses and curved arrays}
Diffraction of the beams as photons propagate between the two mirrors causes small additional loss. As shown in \cite{guimond_subradiant_2019}, this can be mitigated by choosing slightly curved arrays that follow the profile of the optical wavefront. For our Gaussian beams, this gives a near-parabolic surface of the two arrays, such that the phase of the Gaussian field $ {\bf E}_{\rm in} $ is constant across each array. Specifically, atoms at lattice sites $ {\bf r}_n = (x_n,y_n) $ are now placed at $ z = \pm z_{n} $ determined by
\begin{align}
	kz_{n} + k\frac{x_{n}^2 + y_{n}^2}{2R(z_{n})} - \psi(z_{n}) = k\frac{L}{2} - \psi(L/2) .
\end{align}
Such curved arrays improve the coherence properties for large array distances $L$, while diffraction losses generally become insignificant for smaller values of $L\sim\lambda$, which makes it possible to realize narrow transmission resonances and strong photon-photon interactions with flat arrays, compatible with standard optical lattice geometries.   

\subsection{Photon scattering at finite transverse momenta}
\begin{figure}
	\centering
	\includegraphics[width=0.8\textwidth]{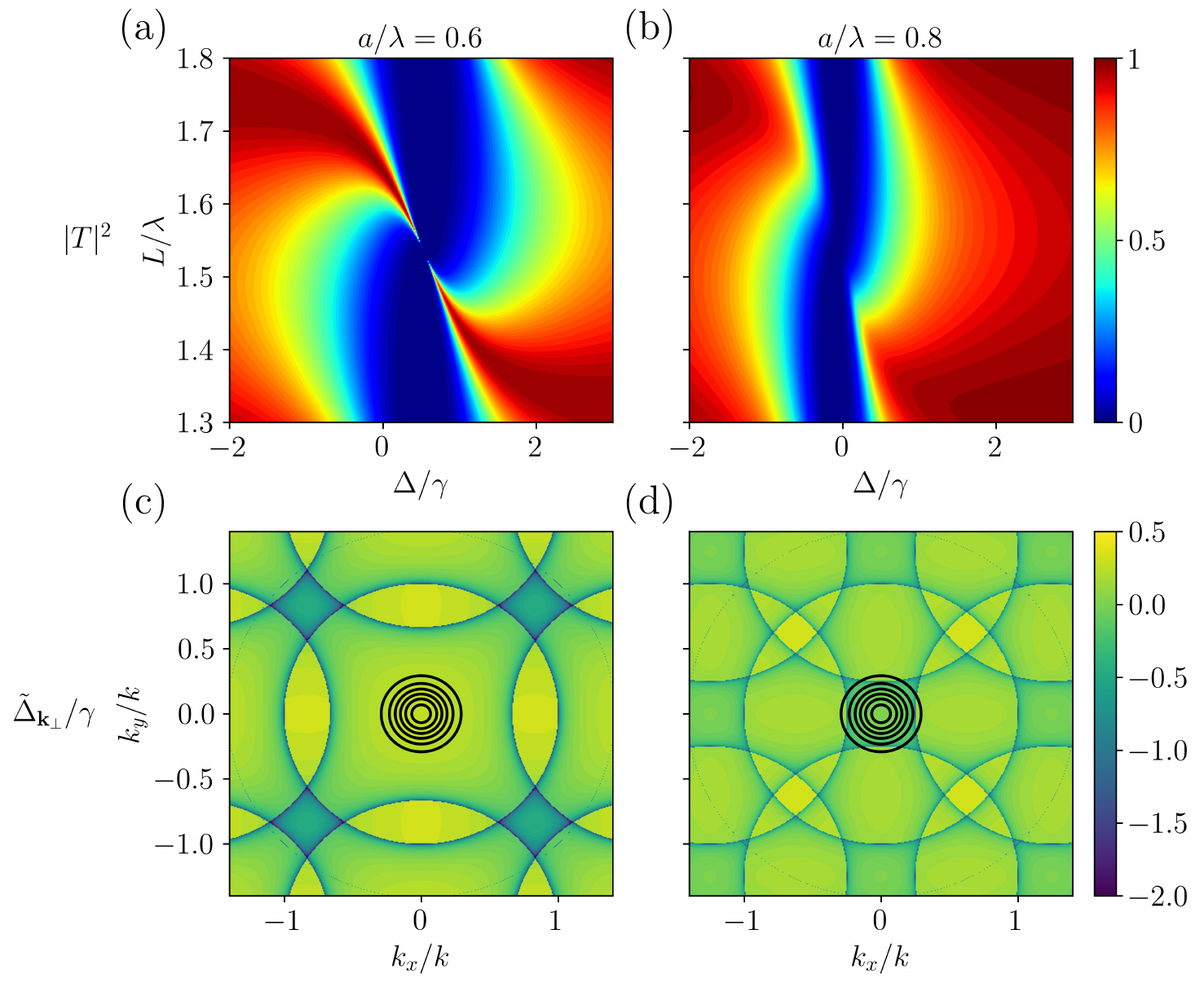}
	\caption{\label{fig4} Transmission spectrum for an infinitely extended dual array but a finite size Gaussian beam with a waist $w=1.5\lambda$ for two different lattice spacings (a) $ a = 0.6\lambda $ and (b) $ a = 0.8\lambda $. Panels (c) and (d) show $ \tilde{\Delta}_{{\bf k}_\perp} $, with values smaller than $ -2\gamma $ cut off, i.e. . The contours of the Gaussian beam profile is indicated by the black circles. The small lattice constant avoids overlap with the resonances of $\tilde{\Delta}_{{\bf k}_\perp}$ and therefore yields much sharper transmission resonances at otherwise identical parameters.}
\end{figure}

Further losses may also arise from the transverse intensity profile of the incident beam, which excites collective atomic excitations at finite transverse momenta with a Rabi frequency $\tilde{\Omega}_{{\bf k}_\perp}$. As a result, there is a finite population of collective excitations with varying shifts $ \tilde{\Delta}_{{\bf k}_\perp} $, such that the resonance condition for perfect transmission cannot be fulfilled for the entire beam. While this remains a minor effect for singlel atomic arrays \cite{manzoni_optimization_2018}, it can have significant consequences for the ultra-narrow resonances of the present dual-array setting.  
As shown in \cref{fig4}b, this effect can indeed alter the optical response in the resonant regime. A closer inspection of the transverse single-array dispersion  $\tilde{\Delta}_{{\bf k}_\perp}$ (cf. \cref{eq:FTG_L}), however, shows that the main broadening stems from sharp resonances of the collective frequency shift. These resonances originate from higher order Bragg scattering involving finite momenta $ {\bf q} = (2\pi/a){\bf m} $. Their effect can thus be suppressed by decreasing the lattice spacing $a$ and eliminated entirely for $a<\lambda/2$. This is illustrated \cref{fig4}, which shows narrow transmission resonances for a beam waist of $w=1.5\lambda$ and a lattice spacing of $a=0.6\lambda$, while they are diminished entirely for $a=0.8$. The former parameters are used for all finite-array calculations discussed in the main text.

\section{Super- and subradiant single-excitation states}
\begin{figure}
	\centering
	\includegraphics[width=0.7\textwidth]{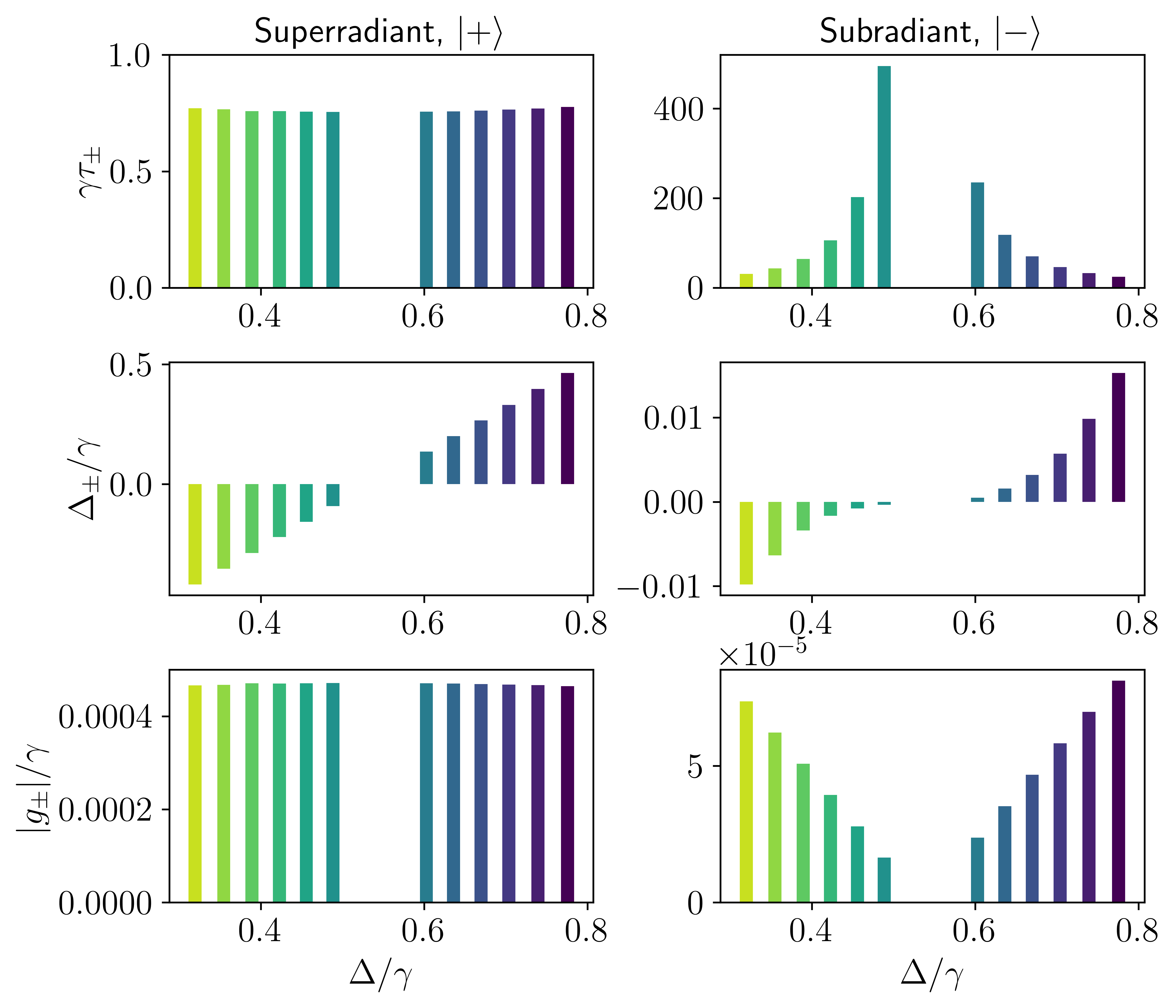}
	\caption{\label{fig5} The detuning $ \Delta_{\pm} $, lifetimes $ \tau_{\pm}=1/\gamma_\pm $, and coupling strengths $ g_{\pm} $ of the two orthogonal collective states $|\pm\rangle$ of the dual array.}
\end{figure}

As discussed in the main text, the characteristic time dependence of $g^{(2)}(t)$ is determined by the dynamics of the short lived and long-lived collective excitation. In the case of infinitely extended arrays and plane wave driving, these states correspond to the symmetric and antisymmetric superposition states $|+\rangle$ and $|-\rangle$, and the two characteristic timescales are set by $\tilde{\Gamma}_\pm$ as introduced above. Following detection of a transmitted photon, the state of the system is projected onto $|\bar{\psi}\rangle = c_{0}|0\rangle + c_{+}|+\rangle + c_{-}|-\rangle $, and the subsequent time evolution of the coefficients can be determined from the non-Hermitian Hamiltonian
\begin{align}
	\begin{split}
		\hat{H} &= -\left(\Delta_{+} + i\gamma_{+}\right)|+\rangle\langle+| - \left(\Delta_{-} + i\gamma_{-}\right)|-\rangle\langle-|\\
			&\hspace{12pt} - g_{+}\left(|+\rangle\langle G| + |G\rangle\langle+|\right) - g_{-}\left(|-\rangle\langle G| + |G\rangle\langle-|\right)
	\end{split}
\end{align}
which coincides with \cref{eq:Heff} above. The solution of the corresponding Schr\"odinger equation gives
\begin{align}\label{eq:sol}
	c_{\pm}(t) = \left(c_{\pm}(0) + \frac{ig_{\pm}}{i\Delta_{\pm} - \gamma_{\pm}}\right)e^{(i\Delta_{\pm} - \gamma_{\pm})t} - \frac{ig_{\pm}}{i\Delta_{\pm} - \gamma_{\pm}}
\end{align}
and yields the dynamics of the probabilities $|c_\pm|^2$ shown in Fig. 2a of the main text. In order to apply this picture to a finite system we numerically simulate its dynamics after photon detection in the absence of the driving field. At long times, this yields the long-lived state following the rapid decay of the superradiant states, which we then construct by orthogonality. Knowing the subradiant and superradiant states, we can then determine the dynamics of $c_\pm(t)$ from the numerical solution of the exact master equation for the finite system, and obtain an excellent fit to \cref{eq:sol}. The extracted effective parameters of the subradiant and superradiant state are presented in \cref{fig5} for the same parameters as in Fig. 3 of the main text. 

\section{Determining the light field}
Having solved for the dynamics of the atomic lattices, the photonic state can be readily reconstructed. The electric field operator of the light field can be obtained from the input-output relation \cite{asenjo-garcia_atom-light_2017}
\begin{align}\label{eq:Eio}
	\hat{{\bf E}}({\bf R}, t) = \hat{\bf E}_{\rm in}({\bf R}, t) + \frac{6\pi \gamma}{k d} \sum_{n}{\bf G}({\bf R}-{\bf R}_{n})\cdot{\bf e}_+\: \hat{\sigma}_{n}(t)
\end{align}
where the incident light is considered to be a coherent field, with a classical amplitude, $ \hat{\bf E}_{\rm in}={\bf E}_{\rm in}$, given by ${\bf E}_{\rm in}=\mathcal{E}_{\rm in}{\bf e}_+f({\bf R})$. Choosing $f({\bf R})$ also as the detection mode, one can define the electric field operator
\begin{equation}
\hat{E}_{f}(t)=\eta^{-1}\int {\rm d}^2 r\: {\bf e}_+^\dagger\cdot\hat{{\bf E}}({\bf R}, t) f^*({\bf R})
\end{equation}
of the transmitted photons ($z\gg L/2$) in the detection mode $f$ with $\int{\rm d}^2r |f({\bf R})|^2=\eta$. Assuming that all fields are paraxial, one then arrives at the simplified form of the input-output relation \cite{manzoni_optimization_2018} given in Eq. (2) of the main text.

Finally, we consider the transverse Fourier transform of the field amplitude in the plane of the atomic arrays 
\begin{equation}
\tilde{E}({\bf k}_\perp)=\int {\rm d}^2 r\: {\bf e}_+^\dagger\cdot\hat{{\bf E}}({\bf R}, t) e^{-i{\bf k}_\perp{\bf r}}e^{-i k_zz},
\end{equation}
as used in Eq. (9) of the main text. The longitudinal wavenumber $k_z=\sqrt{k^2-k_\perp^2}$ is determined by the carrier frequency of the incident field, and we focus here on small transverse momenta $k_\perp<k$ for which $k_z$ remains real. 
The Fourier transformed field is readily obtained from \cref{eq:Eio}
\begin{align}\label{eq:EFT}
	\tilde{E}({\bf k}_\perp, t) &= \mathcal{E}_{\rm in}(t)\tilde{g}({\bf k}_\perp) + \frac{6\pi \gamma}{k d}\sum_{n}\int {\rm d}^2 r\:\: {\bf e}_+^\dagger\!\cdot{\bf G}({\bf R}-{\bf R}_{n})\cdot{\bf e}_+\: e^{-i{\bf k R}}\hat{\sigma}_{n}(t),\nonumber\\
	&= \mathcal{E}_{\rm in}(t)\tilde{g}({\bf k}_\perp) + \frac{6\pi \gamma}{k d} {\bf e}_+^\dagger\!\cdot\tilde{{\bf G}}({\bf k}_\perp,z)\cdot{\bf e}_+\: \tilde{\sigma}_{{\bf k}}(t),\nonumber\\
	&= \mathcal{E}_{\rm in}(t)\tilde{g}({\bf k}_\perp) + i \frac{3\pi \gamma}{2k^3 k_zd} (2k^2-k_\perp^2)\tilde{\sigma}_{{\bf k}}(t),
\end{align}
where $\tilde{\bf G}({\bf k}_\perp,z)=\int {\rm d}^2r\: {\bf G}({\bf R})e^{-i{\bf k R}}$ and $\tilde{\sigma}_{{\bf k}}=\sum_n \hat{\sigma}_{n} e^{-i{\bf k r}_n-ik_z z_n}$  denote the transverse Fourier transforms of the Greens function tensor and the atomic transition operators, respectively. The Fourier transformed transverse mode function $\tilde{g}({\bf k})$ of the incident field is given in \cref{eq:ftilde}. In the last line we have substituted the considered polarization vector $ {\bf e}_{+} = (1,i,0)^{T}/\sqrt{2} $ and assumed $z\gg L/2$, for which the transmitted field, $\tilde{E}({\bf k})$, is independent of $z$, due to the absence of evanescent field contributions. \cref{eq:EFT} has been used to determine the two-photon densities shown in Fig. 4 of the main text.